\documentclass{article}

\usepackage[pagewise]{lineno}
\usepackage[utf8]{inputenc}
\usepackage[margin=1in]{geometry}
\usepackage{graphicx}
\graphicspath{{figs/}}
\usepackage{subcaption}
\usepackage{amsmath, amssymb}
\numberwithin{equation}{section}
\usepackage{xcolor}
\usepackage{natbib}
\usepackage{bm}
\usepackage{nicefrac}
\usepackage{booktabs}
\usepackage{tabularx,tabulary,array}
\newcolumntype{M}[1]{>{\centering\arraybackslash}m{#1}}
\newcolumntype{N}{@{}m{0pt}@{}}
\usepackage{multirow}
\usepackage{dcolumn}
\usepackage{lscape}
\usepackage[normalem]{ulem}
\usepackage[title]{appendix}
\usepackage{authblk}
\usepackage{hyperref}

\newcommand{\lr}[1]{\left(#1\right)}
\newcommand{\vel}{\mathbf{u}} 

\newcommand{\vave}{\overline{\vel}} 
\newcommand{\uave}{\overline{u}} 
\newcommand{\vfluct}{\vel^{\prime}} 
\newcommand{\Rey}{\mathrm{Re}} 
\newcommand{\anirs}{\mathbf{a}} 
\newcommand{\nanirs}{\mathbf{b}} 

\definecolor{HSafron}{RGB}{232,125,30}
\definecolor{HCrimson}{RGB}{165,28,48}

\providecommand{\keywords}[1]
{
  \small	
  \textbf{\textit{Keywords---}} #1
}

\title{Neural Network Models for the Anisotropic Reynolds Stress Tensor in Turbulent Channel Flow}
\author[1]{Rui Fang}
\author[1]{David Sondak}
\author[1]{Pavlos Protopapas}
\author[1,2,3]{Sauro Succi}
\affil[1]{Institute for Applied Computational Science, Harvard University, Cambridge, MA 02138, United States}
\affil[2]{Center for Life Nanoscience at la Sapienza, Istituto Italiano di Tecnologia, viale Regina Elena 295, 00161, Rome, Italy}
\affil[3]{Istituto Applicazioni del Calcolo, National Research Council of Itaty, via dei Taurini 19, 00185, Rome, Italy}
\date{\today}

\begin{document}

\maketitle

\begin{abstract}
Reynolds-averaged Navier-Stokes (RANS) equations are presently one of the most popular models for simulating turbulence. Performing RANS simulation requires additional modeling for the anisotropic Reynolds stress tensor, but traditional Reynolds stress closure models lead to only partially reliable predictions.  Recently, data-driven turbulence models for the Reynolds anisotropy tensor involving novel machine learning techniques have garnered considerable attention and have been rapidly developed.  Focusing on modeling the Reynolds stress closure for the specific case of turbulent channel flow, this paper proposes three modifications to a standard neural network to account for the no-slip boundary condition of the anisotropy tensor, the Reynolds number dependence, and spatial non-locality. The modified models are shown to provide increased predicative accuracy compared to the standard neural network when they are trained and tested on channel flow at different Reynolds numbers. The best performance is yielded by the model combining the boundary condition enforcement and Reynolds number injection. This model also outperforms the Tensor Basis Neural Network~\citep{ling2016reynolds} on the turbulent channel flow dataset. 
\end{abstract}

\keywords{turbulence modeling, Reynolds-averaged Navier-Stokes, deep neural network}

\section{Introduction}\label{sec:intro}
Development of practical, high fidelity turbulence models is a key challenge facing scientists and engineers.  Most fluid systems of interest are in a state of turbulence, which is a disordered fluid flow characterized by many interacting and collectively organized length and time scales.  Predictive models must perform well and be computationally tractable in spite of the challenges posed by the turbulence phenomenon.  The gold standard would be direct numerical simulations (DNS) of turbulent fluid systems, but such simulations will remain largely out of reach at the Reynolds numbers found in nature for the foreseeable future.  A variety of turbulence models have been developed over the decades and have been applied to various engineering fields with differing success rates.  Two dominant approaches in the engineering sciences are Reynolds-averaged Navier-Stokes (RANS) models and Large Eddy Simulation (LES) models, the former being much more computationally tractable than the latter but less accurate.  New hybrid RANS-LES models are under active development to take advantage of the strengths of both approaches~\citep{frohlich2008hybrid}.  In recent years, researchers have started to leverage existing DNS databases of turbulent flows to build machine learning models that learn to represent closures in the RANS equations~\citep{duraisamy2019turbulence}.

The present work focuses on the RANS equations.  The goal in the RANS framework is to determine the average of the flow fields and to model the effects of the fluctuations about the average (called the fluctuating components) on the average fields themselves.  The fluctuating fields enter the RANS equations through the divergence of the Reynolds stress tensor.  It is precisely this tensor that must be modeled.  Traditional Reynolds stress closure models lead to only partially reliable predictions. Two-equation eddy viscosity models are commonly used (e.g. $k-\epsilon$ or $k-\omega$) but they are known to be unable to properly account for streamline curvature and history effects on the individual Reynolds stress components, causing large discrepancies in many engineering-relevant flows~\citep{speziale1991analytical, gatski2004constitutive, johansson2002engineering, chen2003extended}.  Reynolds stress transport equations can also be derived, but these introduce an additional cost as well as require considerable new modeling efforts.  It is expected to be very difficult, if possible at all, to develop a universal representation of the Reynolds stress tensor.  Nevertheless, machine learning approaches can provide substantial flexibility in developing and learning new and more flexible models.

In recent years, machine learning and data science have undergone rapid development thanks to increased data availability, boosted computing power and advanced algorithmic innovations. Many researchers have started to incorporate machine learning and data science techniques into modeling fluid systems~\citep{brunton2016discovering, colabrese2017flow, jimenez2018machine, raissi2019physics}. For RANS turbulence modeling, the task is to learn a good model for the Reynolds anisotropy tensor and considerable effort has been devoted to it~\citep{tracey2015machine, zhang2015machine, ling2016machine, ling2016reynolds, wang2017physics, wu2018physics}. Various supervised learning techniques, including random forests and deep neural networks, have been employed to learn adequate models for the Reynolds anisotropy tensor from high-fidelity DNS data.  In a seminal work,~\citet{ling2016reynolds} utilized a generalized expansion of the Reynolds anisotropy tensor~\citep{pope1975more} and proposed a neural network architecture, called the Tensor Basis Neural Network (TBNN), to learn the coefficients of the expansion. This novel architecture provides an innovative approach to automatically embed the physical invariances in the predicted Reynolds anisotropy tensor. When trained and tested on a variety of flows, the TBNN was shown to provide improved predicative accuracy compared to traditional RANS models as well as a generic neural network architecture that does not embed invariances~\citep{ling2016reynolds}. 

Although significant successes have been achieved, there exist further challenges such as modeling the non-local and non-equilibrium effects of turbulence~\citep{duraisamy2019turbulence}. So far, the vast majority of machine learning models for the Reynolds stress closure, including the TBNN, use only local quantities. However, turbulence is generally a non-local phenomenon in space and time~\citep{speziale1981nonlocal, domaradzki1990local, laval2001nonlocality}.  The Reynolds stress tensor at any location and time can be dependent on prior history of the flow and flow quantities of other spatial locations~\cite{hamlington2009reynolds}.  Spatially non-local effects are particularly significant in strongly inhomogeneous flows, where spatial variations of mean flow quantities are substantial (e.g. the near-wall region of turbulent channel flow)~\citep{hamba2005nonlocal, pope2001turbulent}.  Temporal non-locality effects, also known as non-equilibrium effects, are most obvious when there exist large time variations in the mean strain rate, such as in impulsively and periodically sheared turbulence~\citep{hamlington2008reynolds}. For these flows turbulence models that neglect non-local effects can cause substantial inaccuracies.  It could therefore be of interest to introduce ways to account for non-local effects with the help of machine learning techniques~\citep{song2018universal}.  Interestingly, kinetic formulations of fluid systems can provide insightful perspectives on non-locality~\citep{succi2018lattice}.  Although intriguing, machine learning algorithms applied to kinetic formulations are out of the scope of the present work.

The present work aims to use machine learning to model the Reynolds stress for fully-developed turbulent channel flow.  We use a generic neural network as the base model, and then increase its capabilities by using non-local features, directly incorporating Reynolds-number information, and enforcing the boundary condition at the channel wall.  The models are trained and tested on different combinations of Reynolds numbers in order to investigate the Reynolds-number generalizability. Meanwhile, each of these models is compared to the TBNN. 

The remainder of the paper is structured as follows. Section~\ref{sec:background} provides background on RANS, turbulent channel flow, and neural networks. Section~\ref{sec:models} introduces the new neural network-based models.  Section~\ref{sec:results} describes the datasets used in this work for training the models and presents the Reynolds shear stress predictions by the new models in comparison with the TBNN model for different train-test cases.  Finally, conclusions and future directions are described in Section~\ref{sec:conclusion}.

\section{Background and Methodology} \label{sec:background}
  The focus of this work is the development of new neural network models to represent the Reynolds anisotropy tensor for turbulent channel flow.  The models are tested on turbulent channel flow at a variety of Reynolds numbers.  In this section, we provide some background on turbulent channel flow and neural networks to provide context and to fix notation for the rest of the paper.
  
  \subsection{Turbulent channel flow}
    Turbulent channel flow consists of a fluid confined between two, infinite parallel plates in the $x-z$ plane and situated at $y=0$ and $y=2h$ and the flow is driven by a known pressure gradient in the streamwise ($x$) direction.  The velocity field is denoted by $\vel = (u,v,w)$ where $u$ is the streamwise velocity, $v$ is the wall-normal ($y$) velocity, $w$ is the spanwise ($z$) velocity.  In general, each component of the velocity field is a function of space $(x,y,z)$ and time $t$.  The friction Reynolds number for turbulent channel flow is $Re_{\tau} = u_{\tau}h/\nu$ where $u_{\tau}=\sqrt{\tau_{\text{wall}}/\rho}$ is the friction velocity and $\tau_{\text{wall}}$ is the wall shear stress, which is proportional to the pressure gradient.  The fluid density and kinetic viscosity are denoted by $\rho$ and $\nu$ respectively.  Wall units  are a non-dimensional distance from the wall and are given by $y^+ = u_{\tau}y/\nu$.
    
    The Reynolds-averaged approach to turbulence modeling relies on an averaging operation to decompose the velocity into average and fluctuating components $\vel= \vave + \vfluct$, the goal being to ultimately find the average fields $\vave$.  Performing this averaging operation on the Navier-Stokes equations leads to the Reynolds-averaged Navier-Stokes (RANS) equations and the infamous Reynolds stress tensor $\overline{\vfluct\otimes\vfluct}$, which must be modeled in order to close the RANS equations.  Most modeling efforts actually focus on the anisotropic Reynolds stress tensor $\anirs = \overline{\vfluct\otimes\vfluct} - \lr{2k/3}\mathbf{I}$ where $k$ is the turbulent kinetic energy since that is the portion responsible for turbulent transport. Substantial effort has been expended over the years to understand the nature of and develop useful models for the Reynolds anisotropy tensor.  Eddy viscosity models are among the most popular and widely used models for $\anirs$, at least in part, due to their ease of implementation and quick simulation times.  In particular, the $k-\epsilon$ and $k-\omega$ models have gained widespread acceptance, although they are not universally applicable to all flows of interest.  In recent years, machine learning algorithms have been applied to turbulence datasets to learn the correct Reynolds anisotropy tensor for a variety of flow fields including those where eddy viscosity models are known to be inadequate.
    
    Turbulent channel flow is statistically a one-dimensional flow with mean velocity field $\vave = \lr{\uave\lr{y}, \ 0, \ 0}$.  The RANS equations are to be solved for $\uave$ and only depend on the $u-v$ component of $\anirs$, $a_{uv}$.  Solving for higher moments of the velocity field would require additional equations that would depend on more components of $\anirs$.  
    
    \subsubsection{Turbulent channel flow dataset}
    The dataset used in the present work for training the machine learning models consists of direct numerical simulation (DNS) data at four friction Reynolds numbers $\Rey_{\tau} = \left[550, 1000, 2000, 5200\right]$~\citep{lee2015direct}.  We obtain the data from the Oden Institute turbulence file server\footnote{https://turbulence.oden.utexas.edu}. The DNS in that work was performed with a B-spline collocation method in the wall-normal direction and a Fourier-Galerkin method in the streamwise and spanwise directions. More simulation details can be found in the original reference~\citep{lee2015direct}. The DNS data provide truth labels for the Reynolds anisotropy tensor in this work. The inputs to the machine learning models are mean flow features obtained from RANS simulations of the same flows. In this work, we generated synthetic RANS data by smoothing the DNS fields with a moving average filter of width $3$. The number of available DNS data points varies for different friction Reynolds numbers and therefore the total number of points over the wall-normal direction, $N_y$, also varies.  For $\Rey_\tau = [550, 1000, 2000, 5200]$ the number of points in the wall-normal direction is $N_y = [192, 256, 384, 768]$, respectively.

  \subsection{Neural networks}
    Neural networks are a class of machine learning algorithms that have found applications in a variety of fields, including computer vision~\citep{krizhevsky2012imagenet}, natural language processing~\citep{lecun2015deep}, and gaming~\citep{silver2017mastering}. Neural networks have been shown to be particularly powerful in dealing with high-dimensional data and modeling nonlinear and complex relationships.  Mathematically, a neural network defines a mapping $f: \mathbf{x} \mapsto \mathbf{y}$ where $\mathbf{x}$ is the input variable and $\mathbf{y}$ is the output variable. The function $f$ is defined as a composition of functions, which can be represented through a network structure.  A feed-forward network is generally structured as a set of layers of nodes.  Each layer consists of several nodes (also known as neurons).  The output of layer $i$ is passed to layer $i+1$.  There are no jump connections or feedback loops.  The edges connecting two layers are characterized by a set of unknown weights, $w$, and biases, $b$, so that the output of the previous layer, $h_{i}$, is transformed according to the affine transformation $wh_{i} + b$.  In a fully-connected network every node in the current layer is connected to every node in the previous and next layer.  Each node in the layer applies a nonlinear operation to the affine transformation so that the output of layer $i+1$ is $h_{i+1}=\sigma\lr{wh_{i} + b}$.  The nonlinear function $\sigma$ is usually referred to as the activation function.  It is typically a member of the sigmoid family of functions (such as the logistic function or hyperbolic tangent) although in principle any smooth function should suffice.  The output of the neural network is the predicted function that is parameterized by the weights and biases in the network.  Networks of this form can contain millions of unknown parameters.  The prediction from the neural network is compared to data in a loss function and an optimization algorithm adjusts the weights and biases of the network to minimize this loss function.  The process of optimizing a neural network is a challenge because it is a high-dimensional non-convex optimization problem.  Many optimization strategies exist and new optimization techniques are frequently being proposed~\citep{ruder2016overview}. Commonly used optimization algorithms are gradient-descent based, where the network parameters are iteratively updated according to the gradient of the loss function with respect to the parameters and some learning rate.  For optimizing a neural network, stochastic gradient descent is more often used than classical gradient descent because the latter can easily get stuck in a shallow local minimum~\citep{goodfellow2016deep}. In each iteration of the stochastic gradient descent, the gradient of the loss function on the entire training data is approximated by that on a random batch of training data, thereby stochastic. A full pass over the entire training data is called an epoch.  The fully-connected, feed-forward network (FCFF), also known as the multilayer perceptron (MLP), is only the starting point in a long line of network architectures.  Other network architectures include convolutional networks, recurrent networks, autoencoders, and generative adversarial networks.

    Because of the great success in fields such as image recognition and natural language processing, there is interest in using neural networks in the physical sciences to help model physical processes.  A significant challenge in adopting neural networks for physics-based problems is to inform the network with known physical laws (e.g. conservation laws).  This approach can help facilitate the training process and may provide better generalizability of the learned model to physical parameters regimes in which there is no data.  There exist a variety of approaches in this vein, many of which have been applied to problems in fluid mechanics.  Of particular relevance to the present work are the approaches for learning the Reynolds anisotropy tensor.  One of the first attempts to embed the physical and mathematical structure of the Reynolds anisotropy tensor into a neural network was the Tensor Basis Neural Network~\citep{ling2016reynolds}.  In that work, the authors add an additional tensorial layer to a fully-connected network which returns the most general, local eddy viscosity model~\citep{pope1975more}.  A major benefit of this architecture is that it guarantees Galilean and rotational invariance of the predicted Reynolds anisotropy tensor.
    
    Other possible routes to embedding physics into a neural network are to use physics-inspired activation functions and convolutional layers.  In the current work, we propose three modifications to a fully connected network, which account for some physics of the turbulent channel flow:  1.) Reparameterize the network to enforce $\anirs=\mathbf{0}$ at the boundaries; 2.) Explicitly provide $Re_{\tau}$ as an input to a layer of the network to provide Reynolds number information; 3.) Extension to allow for non-local models.

\section{Models}\label{sec:models}

In this section, we review the TBNN model and propose new neural network architectures to model the Reynolds anisotropy tensor for turbulent channel flow. The new models are developed based on a generic MLP model due to its flexibility. Three modifications are introduced which guarantee $\anirs=\mathbf{0}$ at the boundary, directly incorporate Reynolds number information, and account for non-locality.  At the end of this section, we provide a table summarizing the characteristics of the various models. 

  \subsection{Review of TBNN}
    
    Developed by~\citet{ling2016reynolds}, the Tensor Basis Neural Network (TBNN) is a novel neural network architecture that embeds physical invariance properties into the modeled Reynolds anisotropy tensor. The core innovation of the TBNN is the design of a network architecture that represents a general eddy viscosity model~\citep{pope1975more}, which by itself guarantees the symmetry and invariance properties of the predicted anisotropy tensor while accounting for physics beyond that permitted by a linear eddy viscosity model. 
    
    The general eddy viscosity model refers to the most general representation of the Reynolds anisotropy tensor in terms of the mean velocity gradient, which is~\citep{pope1975more}
    \begin{align}
      \nanirs = \sum_{n=1}^{10}g^{(n)}\left(\lambda_1, ..., \lambda_5\right) \mathbf{T}^{(n)},
      \label{eq:nlev}
    \end{align}
    where $\nanirs$ is the Reynolds anisotropy tensor normalized by $2k$, $\mathbf{T}^{(n)}$ are basis tensors and $g^{(n)}\left(\lambda_1, ..., \lambda_5\right)$ are coefficients that depend on the five scalar tensor invariants. The basis tensors $\mathbf{T}^{(n)}$ and the scalar invariants $\lambda_m$ are known functions of the symmetric and anti-symmetric part of the normalized mean velocity gradient tensor. Finding the explicit expression of the coefficients is extremely difficult for general three-dimensional turbulent flows, with the significant aggravation that there is no obvious hierarchy of the basis components.  The approach taken by~\citet{ling2016reynolds} was to train a deep neural network to learn the coefficients and subsequently the Reynolds anisotropy tensor across a variety of flow fields. 

\begin{figure}[htp]
    	\centering
    	\includegraphics[width=0.8\columnwidth]{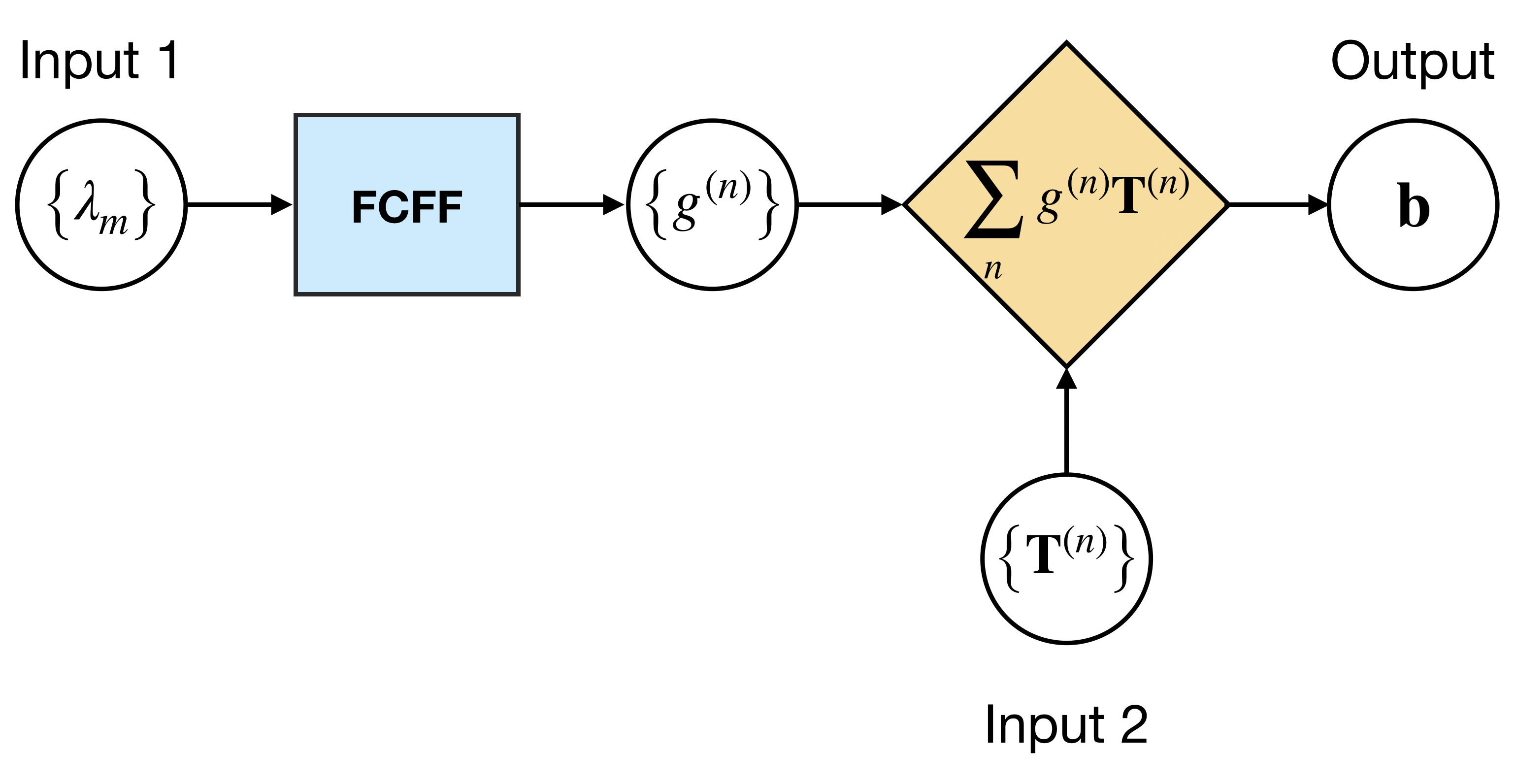}
        \caption{Diagram of the TBNN.  The invariants are passed to the fully-connected feed-forward (FCFF) network, which outputs the tensor basis coefficients.  A second input layer provides the actual tensor basis, which is combined with the predictions from the FCFF network to provide the normalized anisotropy tensor.}
        \label{fig:tbnn}
    \end{figure}

    A schematic of the TBNN is provided in Figure~\ref{fig:tbnn}. The TBNN consists of two input layers.  The first input layer is formed by the scalar invariants, which are passed to a fully-connected feed-forward (FCFF) network.  The second input layer is the basis tensors.  They are combined with the ten tensor basis coefficients predicted from the FCFF to form the normalized anisotropy tensor $\nanirs$, according to~\eqref{eq:nlev}. 
    
    This architecture guarantees that $\mathbf{b}$ will satisfy invariance properties and remain a symmetric, anisotropic tensor. The RANS equations are known to be invariant to rotations and reflections of coordinate axes and translations of the frame of reference with constant speed (also known as Galilean invariance). Therefore, it is important for any machine learning approach to preserve these invariance properties in the predicted flow variables. In addition, embedding physics into the machine learning model provides benefits to the model performance. As demonstrated by the authors in the original paper, when trained and tested on a variety of flows, the TBNN yielded the best predictions of the Reynolds anisotropy tensor compared to traditional RANS models and a generic neural network that does not embed invariance properties. 
    
    On the other hand, the TBNN still faces challenges inherent to the general eddy viscosity model. For example, the general eddy viscosity model represents the anisotropy tensor as a function of local mean velocity gradient, and has no expression of non-locality. It is of our interest to design new network architectures that are competitive with the TBNN while being more flexible to address its shortcomings.


  \subsection{Base model - MLP}

\begin{figure}[htp]
    	\centering
    	\includegraphics[width=0.5\columnwidth]{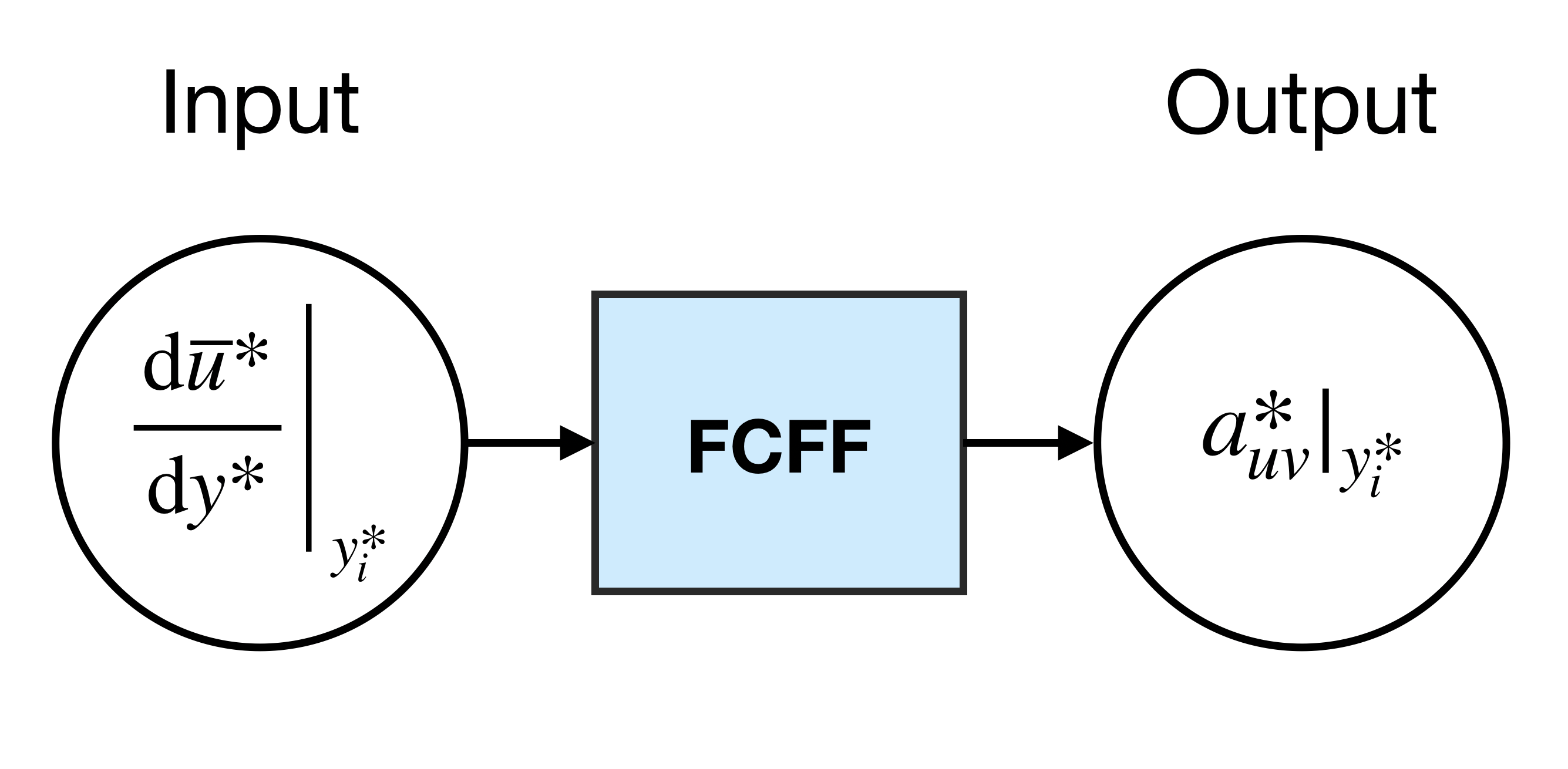}
        \caption[Diagram of the MLP. ]{Diagram of the MLP for turbulent channel flow.  The input is the non-dimensionalized velocity gradient evaluated at a specific point.  The output is the predicted anisotropy tensor for turbulent channel flow.}
        \label{fig:MLP}
    \end{figure}

    A generic MLP model is employed to serve as the simple baseline neural network model. Figure~\ref{fig:MLP} shows the diagram of the MLP.  Starred quantities denote non-dimensionalized quantities.  In particular, the spatial dimension $y$ is normalized by the half channel width $h$: $y^*=y/h$. The mean velocity and the Reynolds anisotropy tensor are normalized by the bulk velocity $u_b = \frac{1}{h}\int_{0}^h \uave \mathrm{d}y$: $\uave^* = \uave/u_b$, $\overline{u^\prime v^\prime}^* = \overline{u^\prime v^\prime}/u_{b}^2$.  The input to the model is the mean velocity gradient of a channel flow $\frac{\mathrm{d}\uave^*}{\mathrm{d}y^*}$ and the output of the model is the $u-v$ component of the Reynolds anisotropy tensor $a_{uv}^*$.  The input is fed through a fully-connected feed-forward network, which produces the output. To emphasize that this is a local model, the input and output are specified at a particular location $y^*_i$.
    
    We note that the non-dimensionalization employed here as well as the predicted output are specific to RANS for turbulent channel flow, which is the focus of this paper.  The RANS equation for $\uave$ only depend on $a_{uv}$ and therefore prediction of the $u-v$ component of the anisotropy tensor is sufficient to solve for the average velocity field.  This model can be generalized in a straightforward way to predict the $6$ components of the anisotropy tensor for more involved flow fields.  A more general non-dimensionalization would involve the prediction of the normalized anisotropy tensor following the TBNN model.
    
    Compared with the TBNN, the MLP is rather simple and has no embedded physics other than Galilean invariance by using the velocity gradients as inputs. Nevertheless, we choose to use this model for its flexibility.

  \subsection{Boundary condition enforcement}

\begin{figure}[htp]
    	\centering
    	\includegraphics[width=0.5\columnwidth]{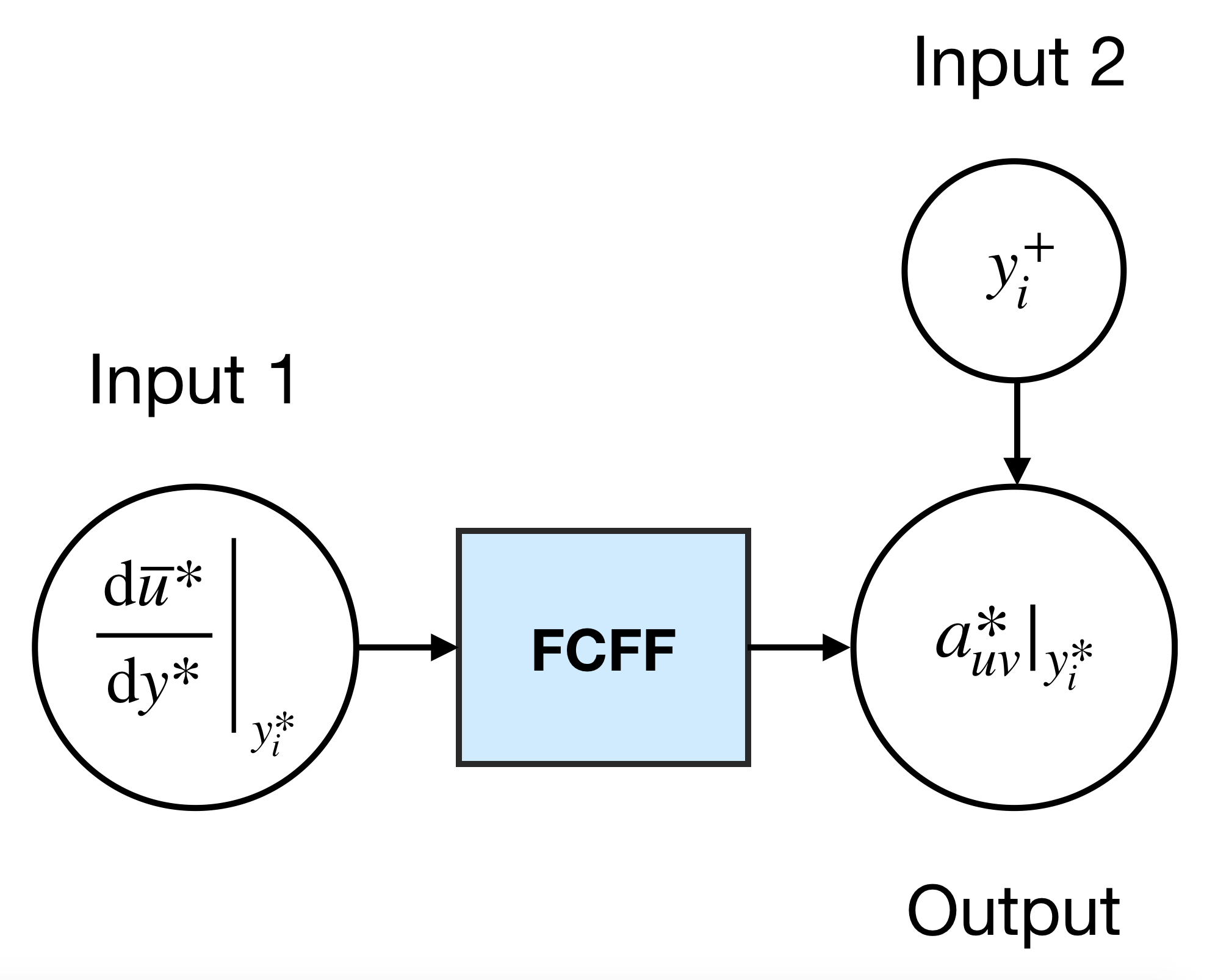}
        \caption[Diagram of the MLP with boundary condition enforcement. ]{Diagram of the MLP with boundary condition enforcement. The boundary condition is enforced through multiplying the FCFF output by a user-selected function of $y^+$ with the property of being zero at $y^+=0$.}
        \label{fig:MLP+BC}
    \end{figure}

    The first improvement that can be done to the MLP is enforcing the no-slip boundary condition at the channel wall. The motivation for this is that the MLP is a free form with almost no embedded physics, hence it is easy to lead to erroneous and unrealistic predictions. To impose the boundary condition, we reparameterize the solution as follows,
    \begin{equation}
        a_{uv}^*\lr{\frac{\mathrm{d}\uave^*}{\mathrm{d}y^*}, y^+} =  A\lr{y^+}\mathrm{FCFF}\lr{\frac{\mathrm{d}\uave^*}{\mathrm{d}y^*}}
    \end{equation}
    where $\mathrm{FCFF}\lr{\frac{\mathrm{d}\uave^*}{\mathrm{d}y^*}}$ is the output of an MLP, $y^+$ is the distance from the wall in viscous units, and $A\lr{y^+}$ is a user-selected function with the property that $A\lr{0} = 0$. We choose $A\lr{y^{+}} = 1-e^{-\beta y^+}$ with $\beta$ a hyperparameter controlling the shape of the function. 
    In this way, the solution satisfies the boundary condition $a_{uv}^*=0$ at $y^+=0$ by construction. This is illustrated in Figure~\ref{fig:MLP+BC}. Note that the model takes $y^+$ as an extra input and the loss function is calculated based on the reparameterized solution. We expect the correct boundary condition to help improve the overall predictive accuracy as measured by the $R^2$ score. 

  \subsection{Reynolds number injection}
    
    The Reynolds number of a flow greatly influences the mean velocity profile. In turbulence modeling with machine learning approaches, it is a natural practice to include the Reynolds number as a flow feature. For example~\citet{wang2017physics} used the wall-distance based Reynolds number as an indicator to distinguish boundary layers from shear flows in their random forest models. Note that we also use a similar measure, $y^+$, in the boundary condition enforcement. However, $y^+$ only carries indirect information of the Reynolds number. We expect a more direct injection of the Reynolds number would affect the model differently. 
    
\begin{figure}
        \centering
        \begin{subfigure}[b]{0.53\columnwidth}
            \includegraphics[width=1.0\columnwidth]{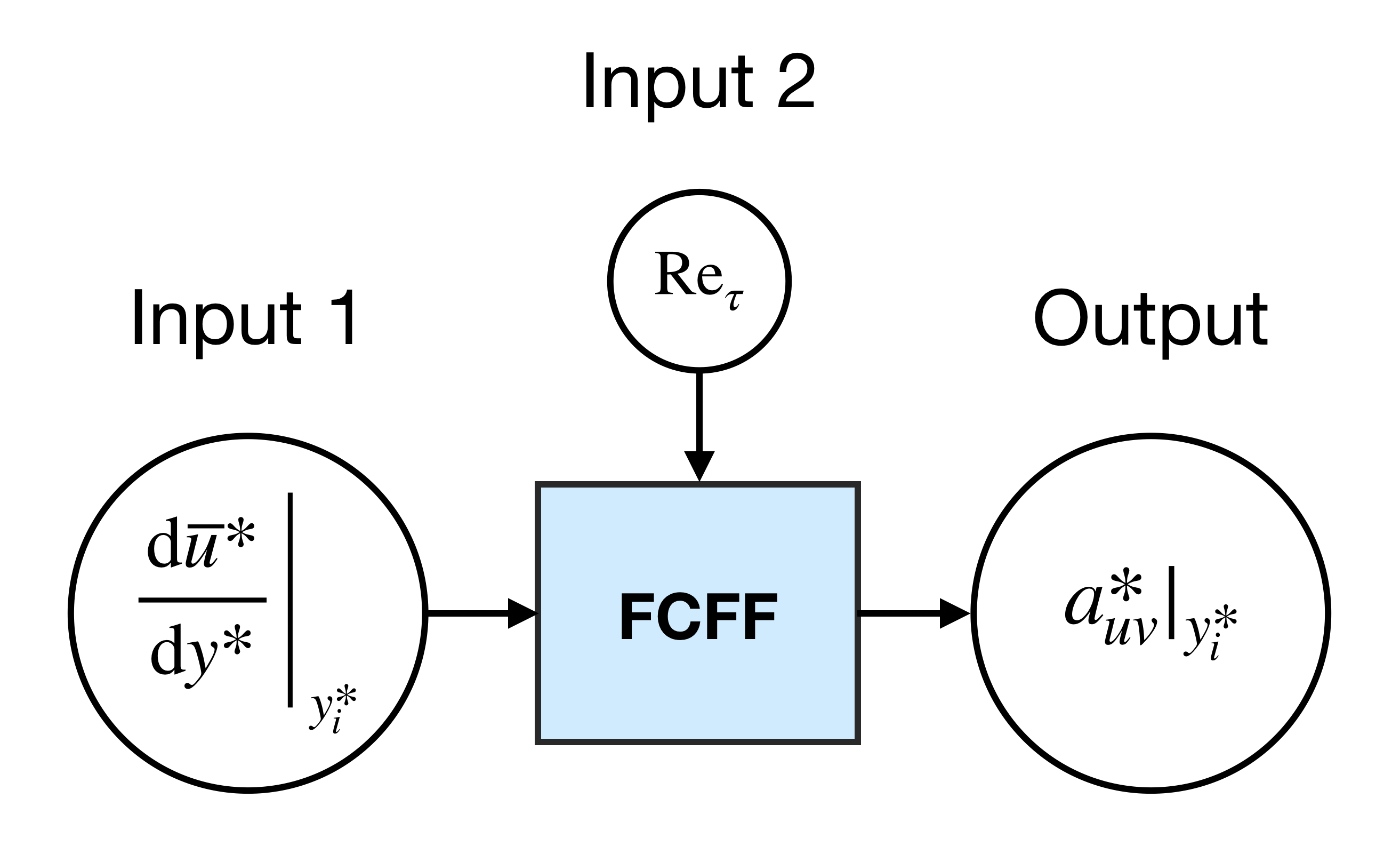}
            \caption{}
            \label{fig:MLP+Retau}
        \end{subfigure}
        
        \begin{subfigure}[b]{0.7\columnwidth}
            \includegraphics[width=1.0\columnwidth]{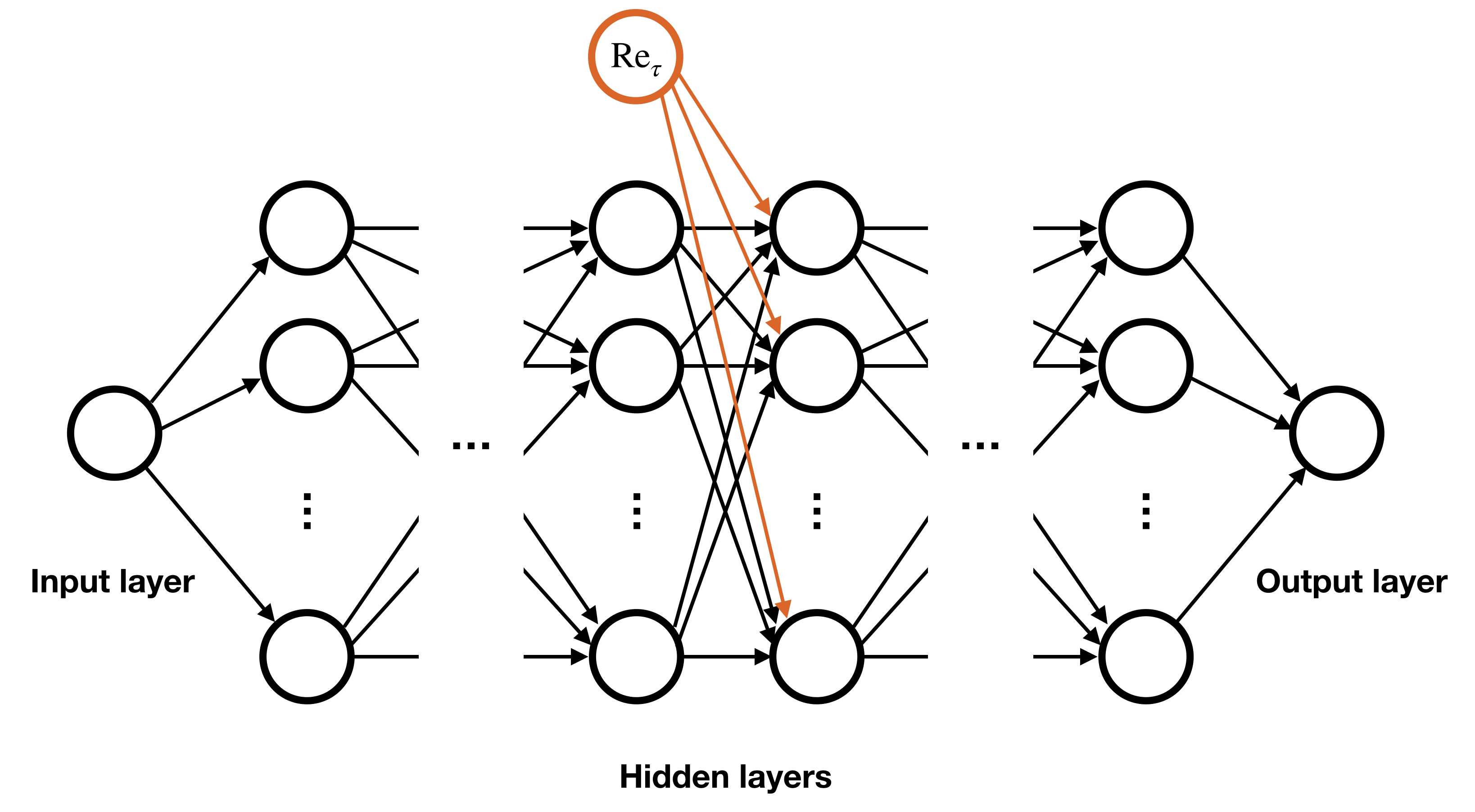}
            \caption{}
            \label{fig:MLP+Retau_detail}
        \end{subfigure}
        \caption{Diagrams of the MLP with additional $\Rey_\tau$ input. (a) Reynolds number information is injected into the FCFF network at one or more layers after the input. (b)  A more detailed depiction of how the Reynolds number is input into each node of a single hidden layer in the FCFF network.}
    \end{figure}

    Figure~\ref{fig:MLP+Retau} provides the diagram of a modified MLP with an additional input $\Rey_\tau$, the friction Reynolds number for the channel flow. Figure~\ref{fig:MLP+Retau_detail} illustrates in detail how the $\Rey_\tau$ is given to the network. It is typically fed into one or more of the intermediate layers of a fully-connected feed-forward network. The intuition is to give the network some higher-level information, which needs to be separated from the ordinary input. 

  \subsection{Non-locality}

    Over the years, many efforts have been devoted to improving turbulence models to better account for non-local effects. Some are shown to give promising results for the case of turbulent channel flow. ~\citet{hamlington2009nonlocal} developed a classical non-local turbulence model by using a series of Laplacians of the mean strain rate tensor to express the rapid pressure-strain correlation. When applied to turbulent channel flow, their closure resulted in good agreement with the DNS values~\citep{hamlington2009nonlocal}. More recently, a group of researchers employed fractional calculus to model the spatial non-locality in wall-bounded turbulent flows~\citep{song2018universal}. They replaced the original RANS equation with a variable-order fractional differential equation to model the Reynolds stresses. By fitting the DNS data for turbulent channel flow at several different Reynolds numbers, they found a universal form for the variable fractional order, which also holds for turbulent Couette flow and pipe flow. 
    
\begin{figure}[htp]
    	\centering
    	\includegraphics[width=1.0\columnwidth]{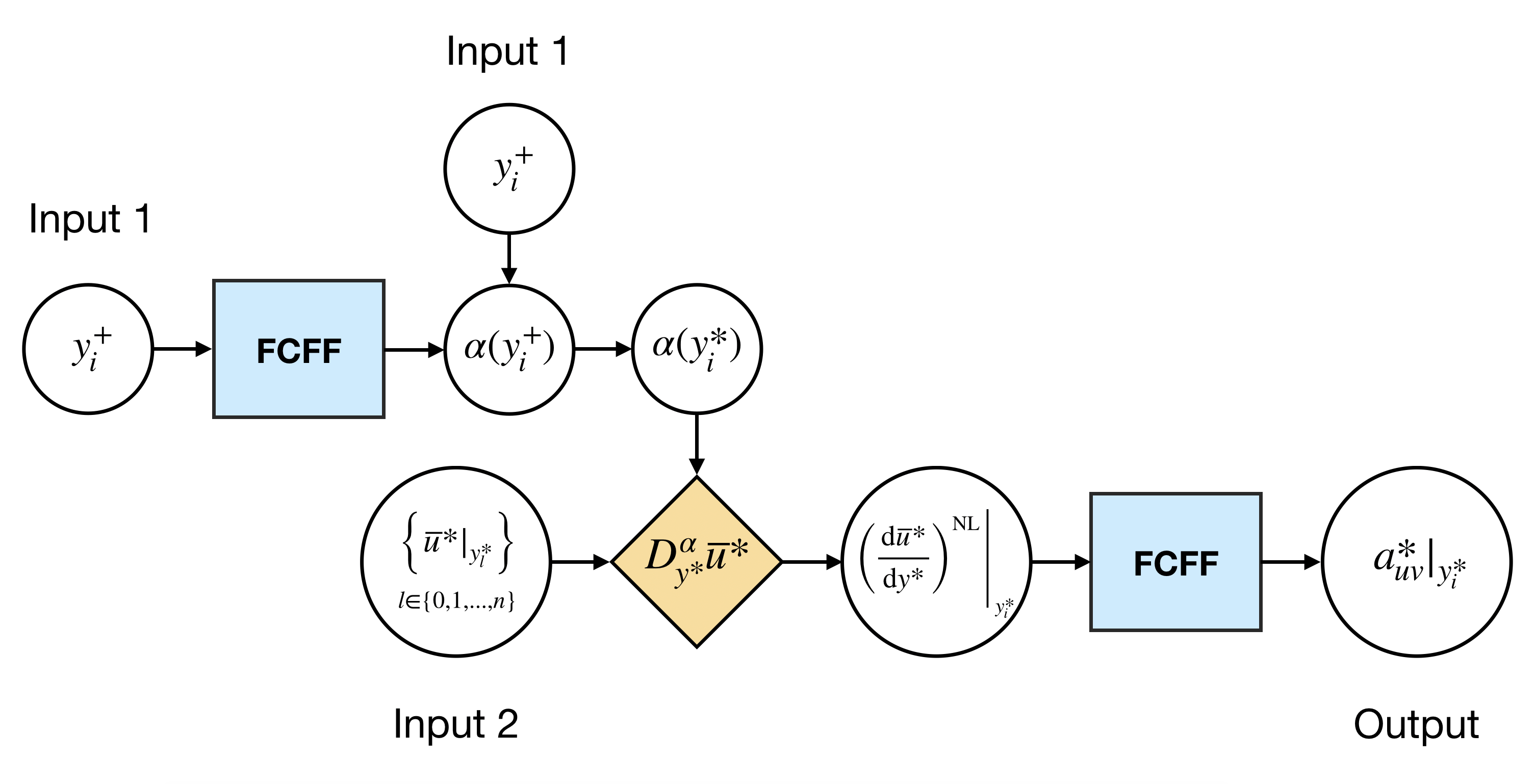}
        \caption{Diagram of the MLP with non-local feature.  This architecture consists of two FCFF networks.  The first network outputs a variable, fractional differential operator order $\alpha$.  Then $\alpha$ is used in a deterministic step to compute the non-local Caputo derivative of the velocity field representing a non-local velocity gradient.  This gradient is the input to a second FCFF network the output of which is the predicted non-local anisotropy tensor.}
        \label{fig:MLP+NL}
    \end{figure}

    We design the following modification to the MLP to capture non-local effects in turbulent channel flows. The network architecture is shown in Figure~\ref{fig:MLP+NL}.  Like the previous architectures, this architecture includes a FCFF network that takes as input a velocity gradient and predicts the anisotropy tensor (the second FCFF network on the right in Figure~\ref{fig:MLP+NL}).  However, the input to this FCFF network is a non-local velocity gradient, $\frac{\mathrm{d}\uave^*}{\mathrm{d}y^*}$, which is a fractional derivative of $\uave^*$ with variable fractional order $\alpha\lr{y^+}$.  We therefore refer to the output as the non-local anisotropy tensor.  A major new addition is that the order of the non-local velocity gradient is learned from a different FCFF network.  This process is represented by the first FCFF network on the left in Figure~\ref{fig:MLP+NL}.
    The input to the first FCFF network is $y^+_i$ associated with some particular location $y^*_i$, and this is used to predict the corresponding fractional order $\alpha$ (assuming $0\leq\alpha\leq1$); the second input is $\uave^*$ at all locations $y^*_l, l\in\{0,1,...,n\}$.  The non-local derivative for the location $y^*_i$ is calculated using the Caputo fractional derivative~\citep{li2015numerical},
    \begin{align}
        \left.\lr{\frac{\mathrm{d}\uave^*}{\mathrm{d}y^*}}^{\mathrm{NL}} \right|_{y^*_i}
        &= \left. D_{y^*}^{\alpha\lr{y^*}} \uave^*\lr{y^*}\right|_{y^*_i} 
        = \frac{1}{\Gamma\lr{1-\alpha\lr{y^*_i}}} \int_0^{y^*_i} (y^*_i-\tau)^{-\alpha\lr{y^*_i}} \frac{\mathrm{d}\uave^*}{\mathrm{d}\tau} \mathrm{d}\tau 
        \label{eq:frac_deriv}
    \end{align}
    
    It can be proved that, for $y^*_i\neq0$,~\eqref{eq:frac_deriv} becomes $\uave^*\rvert_{y^*_i}$ when $\alpha\lr{y^*_i}=0$, and becomes $\frac{\mathrm{d}\uave^*}{\mathrm{d}y^*}\rvert_{y^*_i}$ when $\alpha\lr{y^*_i}=1$~\citep{li2015numerical}. In other words,~\eqref{eq:frac_deriv} reduces to local flow quantities when $\alpha$ is an integer. 
    At the lower terminal of integration ($y^*_i=0$),~\eqref{eq:frac_deriv} is undefined unless $\alpha\lr{0}$ is 0 or 1. If $\alpha\lr{0}=0$, then~\eqref{eq:frac_deriv} is $\uave^*\rvert_{y^*_i=0}$; if $\alpha\lr{0}=1$, then~\eqref{eq:frac_deriv} is $\frac{\mathrm{d}\uave^*}{\mathrm{d}y^*}\rvert_{y^*_i=0}$. To be consistent with the baseline MLP model and traditional RANS models, where the Reynolds anisotropy tensor at the wall depends on the velocity gradient, we decide to let $\alpha\lr{0}=1$. This condition is enforced by reparameterizing $\alpha\lr{y^+}$: 
    \begin{equation}
        \alpha\lr{y^+} = 1 - \lr{1-e^{-\beta y^+}}\mathrm{FCFF}\lr{y^+}.
    \end{equation}
    Here $\mathrm{FCFF}\lr{y^+}$ represents the output of a fully-connected feed-forward network, whose output activation function is set to be a sigmoid function so that the returned value is inside $[0, 1]$. Multiplying it by $\lr{1-e^{-\beta y^+}}$ and then subtracting the product from 1 forces $\alpha=1$ at $y^+=0$. Because $0<1-e^{-\beta y^+}<1$ for $y^+>0$, the resulting $\alpha\lr{y^+}$ is always inside $[0, 1]$ as desired.

    Overall, this non-local modification has several advantages. First, since the fractional derivative is an integral when $\alpha$ is not an integer, it is able to express non-locality. Second, the fractional order $\alpha$ is itself a function of space, allowing the encoding of variable ``strength'' of the non-locality across space. Last, through fitting the function $\alpha\lr{y^+}$, the model is able to learn universality among channel flows with different $\Rey_\tau$. That is, the non-local parameter $\alpha$ will reflect the fact that mean flow profiles in turbulent channel flow collapse to the same curve when plotted in $+$ units over a range of $Re_{\tau}$~\citep{song2018universal, pope2001turbulent}.

  \subsection{Combination of models}
    
    The above modifications can be combined to develop more complex architectures with more capabilities. For instance, it is easy to combine boundary condition enforcement and Reynolds number injection in a single model. Based on this model, we will explore two types of combination:  with and without the non-locality. Table~\ref{tab:models} summarizes the capabilities of all proposed models and the TBNN model. 
    
   \begin{table}[htp]
        \centering
        \caption{Comparison of various models.}
        \begin{tabular}{ccccc}
            \toprule
                 & no-slip B. C. & spatial non-locality  & Galilean invariance & rotational invariance \\
            \midrule
            MLP &  &  & \checkmark & \\ 
            MLP-BC & \checkmark &  & \checkmark & \\
            MLP-$\Rey_\tau$ & & & \checkmark & \\
            MLP-NL & &  \checkmark & \checkmark & \\
            MLP-BC-$\Rey_\tau$ & \checkmark & & \checkmark & \\
            MLP-BC-$\Rey_\tau$-NL & \checkmark &  \checkmark  & \checkmark & \\
            TBNN & \checkmark & & \checkmark & \checkmark \\
            \bottomrule
        \end{tabular}
        \label{tab:models}
    \end{table}

\section{Results}\label{sec:results}

\begin{figure}[htp]
    \centering
    \includegraphics[width=0.7\linewidth]{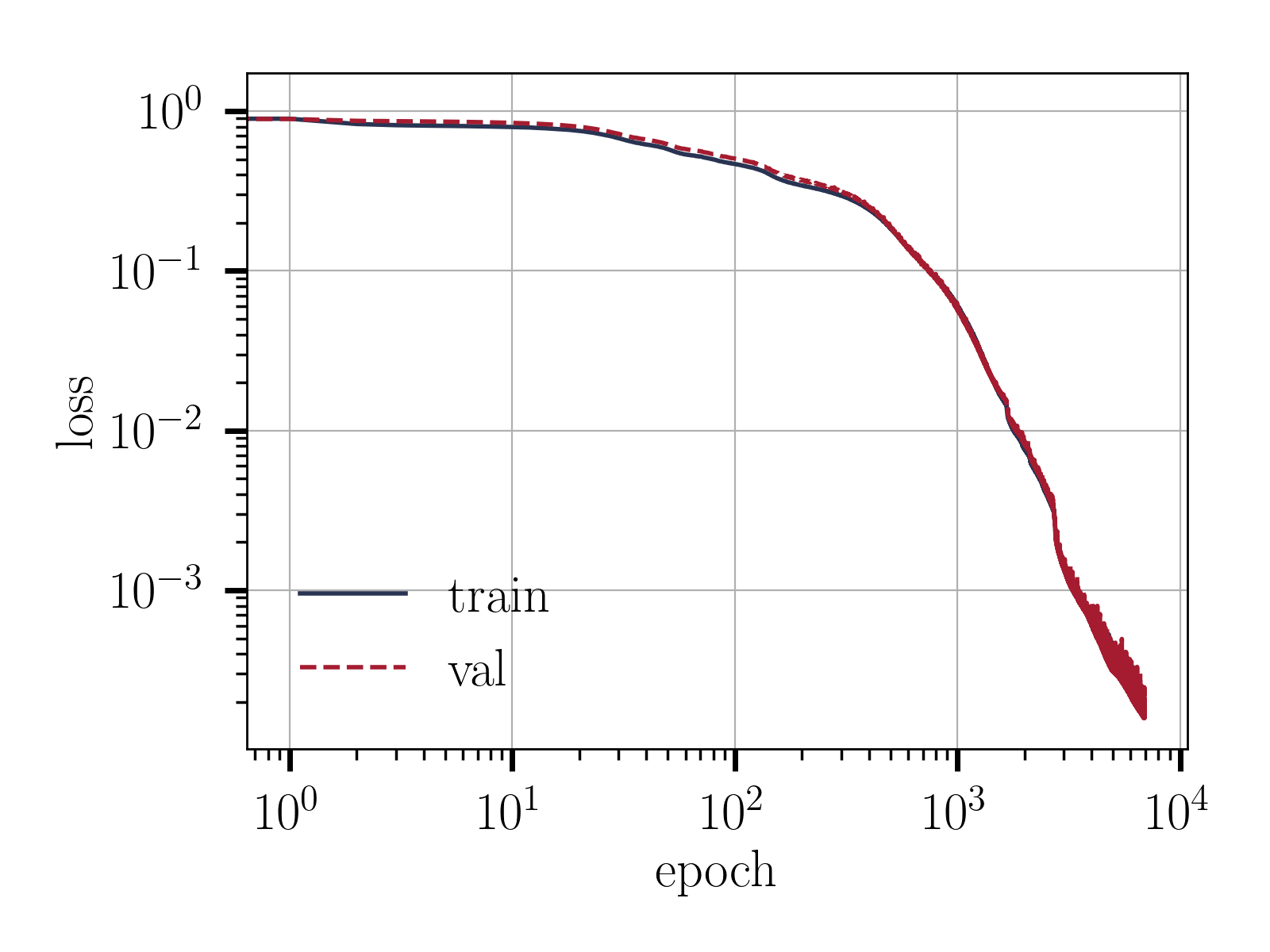}
    \caption{The training and validation loss as a function of epochs during training the MLP-BC-$\Rey_\tau$ model for Case 1. The model has converged after approximately $7000$ epochs.}
    \label{fig:training}
\end{figure}

Table~\ref{tab:train_test_cases} displays the four training-prediction cases that are examined. For each case, we randomly split the training data into 80\% training and 20\% validation data.  The loss function was the mean squared error between the DNS and the predicted values of $a_{uv}^*$. The Adam optimization algorithm~\citep{kingma2014adam} with an initial learning rate of $10^{-6}$ and a batch-size of 10 was used to train the networks. The weights and biases of the networks were initialized using the He uniform distribution~\citep{he2015delving}. We used early-stopping~\citep{prechelt1998early} as a regularization technique to guard against over-fitting. Early-stopping refers to the procedure where the validation loss is monitored during training and the training process terminates once the validation loss begins to increase. In machine learning, an increase in validation loss is usually a strong indicator of over-fitting~\citep{goodfellow2016deep}. Figure~\ref{fig:training} shows an example of the training and validation loss as a function of epochs during training the MLP-BC-$\Rey_\tau$ model for Case 1.  In this case, the model was trained after $6924$ epochs.  From the figure, it appears that the loss has not yet converged.  This is because the early-stopping criterion was used.  Shortly after epoch $6924$, the validation loss starts to rise relative to the training loss.  The early stopping criterion senses this and terminates the training process.  The number of training epochs varied for different models and different cases, but was generally in the range of 5000-10000.

For model evaluation, we used the $R^2$ score, a statistical measure of how well the regression predictions approximate the true data. An $R^2$ of 1 indicates a perfect fit. The $R^2$ scores of the various models on the training and test data are displayed in Table~\ref{tab:model_performance_r2}. The $R^2$ scores on the validation data are very similar to the training $R^2$ and are therefore omitted in the reported results. 
In subsequent sections we discuss the performance of each of the MLP-BC, MLP-$\Rey_\tau$, MLP-NL, and the combined models, with comparisons to the base MLP model and the TBNN model.


    \begin{table}[htp]
        \centering
        \caption{Four training-prediction cases.}
        \begin{tabular}{ccc}
            \toprule
                Case & Training set & Test set \\
            \midrule
            1 & $\Rey_\tau=[550, 1000, 2000]$  & $\Rey_\tau=5200$ \\ 
            2 & $\Rey_\tau=[550, 1000, 5200]$  & $\Rey_\tau=2000$ \\
            3 & $\Rey_\tau=[550, 2000, 5200]$  & $\Rey_\tau=1000$ \\
            4 & $\Rey_\tau=[1000, 2000, 5200]$ & $\Rey_\tau=550$  \\
            \bottomrule
        \end{tabular}
        \label{tab:train_test_cases}
    \end{table}

    \begin{table}[htp]
        \caption{$R^2$ of $a_{uv}^*$ predictions by various models for four training-prediction cases.}
        \begin{tabular}{@{}lllllllll@{}}
            \toprule
             & \multicolumn{2}{c}{Case 1} & \multicolumn{2}{c}{Case 2} & \multicolumn{2}{c}{Case 3} & \multicolumn{2}{c}{Case 4} \\
             \cmidrule(l){2-3} \cmidrule(l){4-5} \cmidrule(l){6-7} \cmidrule(l){8-9} 
             & Train $R^2$ & Test $R^2$ & Train $R^2$ & Test $R^2$ & Train $R^2$ &  Test $R^2$ & Train $R^2$ & Test $R^2$ \\
            \midrule
            MLP-BC-$\Rey_\tau$-NL & 0.9998 & 0.8196 & 0.9999 & 0.9959 & 0.9999 & 0.9996 & 0.9998 & 0.9620 \\
            MLP-BC-$\Rey_\tau$    & 0.9997 & 0.9628 & 0.9991 & 0.9970 & 0.9996 & 0.9938 & 0.9988 & 0.9783 \\
            MLP-$\Rey_\tau$       & 0.9997 & 0.9476 & 0.9997 & 0.9959 & 0.9996 & 0.9893 & 0.9987 & 0.9479 \\
            MLP-NL                & 0.9744 & 0.7683 & 0.9377 & 0.9705 & 0.9442 & 0.9512 & 0.9675 & 0.8274 \\
            MLP-BC                & 0.9499 & 0.5441 & 0.8672 & 0.9880 & 0.8990 & 0.8829 & 0.9355 & 0.7239 \\
            MLP                   & 0.9238 & 0.5186 & 0.8301 & 0.9390 & 0.8470 & 0.8531 & 0.9108 & 0.5885 \\
            TBNN                  & 0.9431 & 0.9397 & 0.9493 & 0.9632 & 0.9640 & 0.9588 & 0.9582 & 0.8983 \\ 
            \bottomrule
        \end{tabular}
        \label{tab:model_performance_r2}
    \end{table}

  \subsection{Performance of the MLP-BC model}

    First, we investigate the performance of the MLP-BC model by comparing it to the base MLP. The base MLP has 5 hidden layers with 50 nodes per layer. The nonlinear activation function of the hidden nodes is the Exponential Linear Unit (ELU)~\citep{clevert2015fast}.  The FCFF network in the MLP-BC model shares the same structure as the MLP. The hyperparameter $\beta$ is chosen to be $0.1$. 
    
\begin{figure}[htp]
        \centering
        \begin{subfigure}{0.49\textwidth}
            \centering
            \includegraphics[width=1.0\columnwidth]{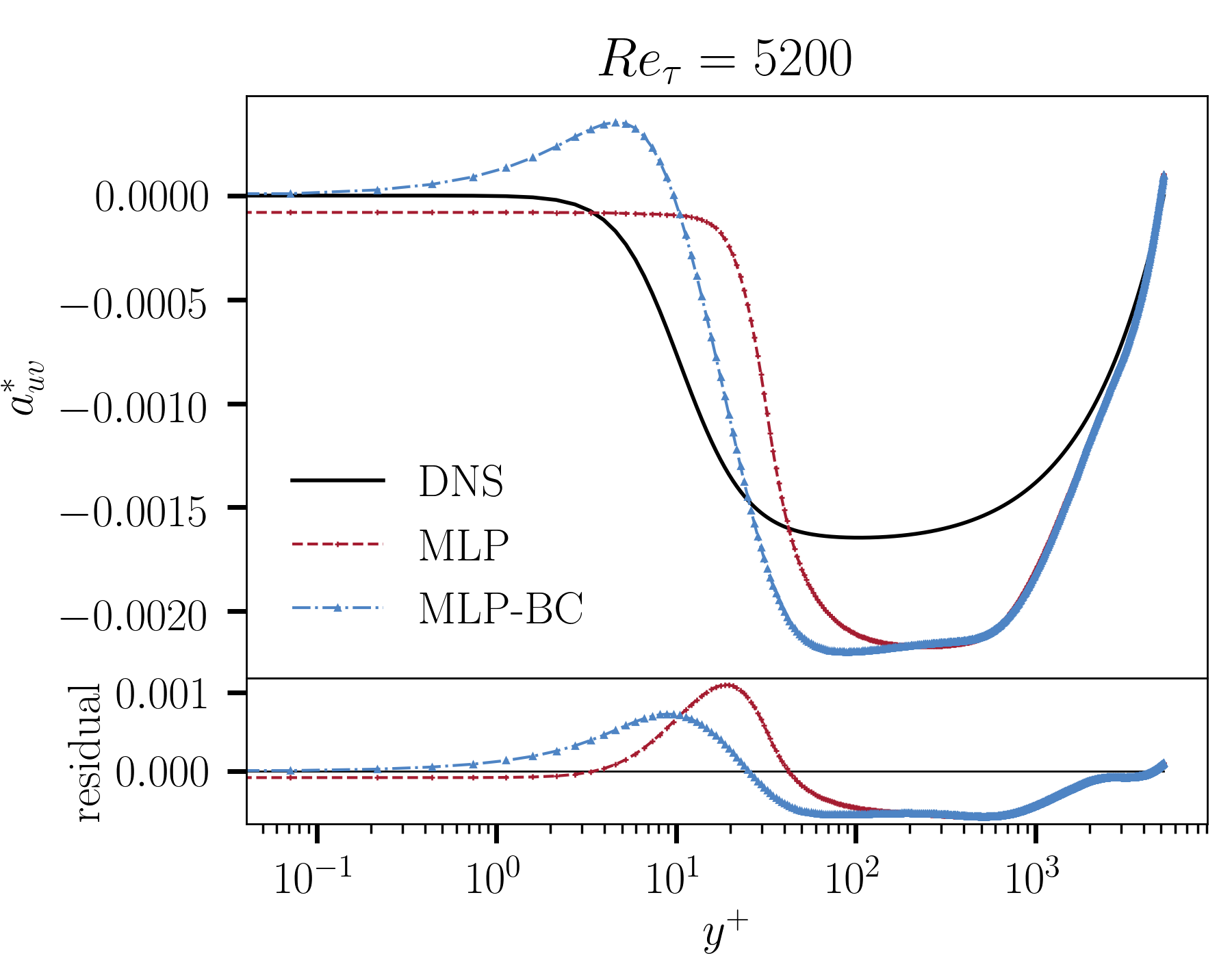}
            \caption{}
        \end{subfigure}
        ~
        \begin{subfigure}{0.49\textwidth}
            \centering
            \includegraphics[width=1.0\columnwidth]{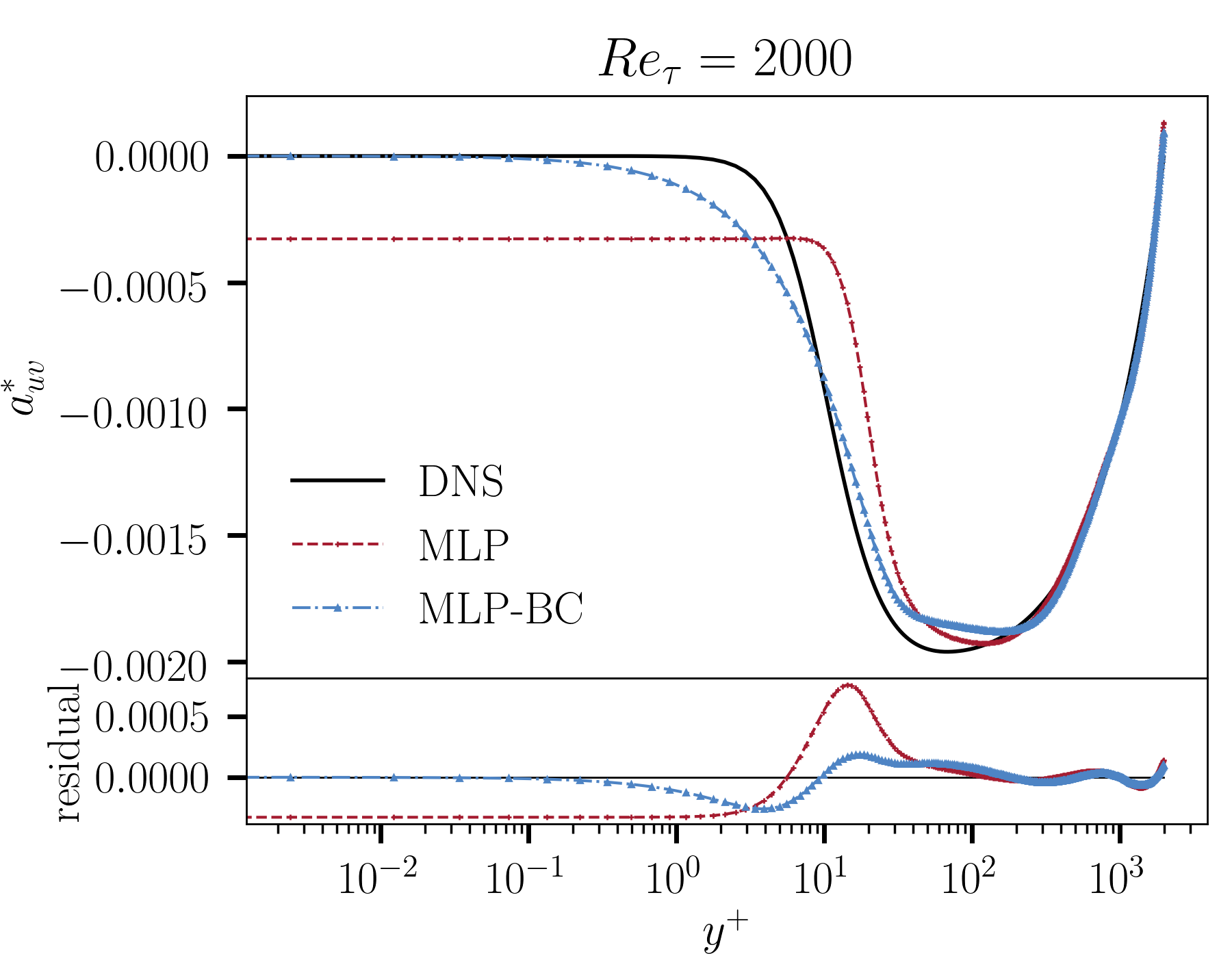}
            \caption{}
        \end{subfigure}
        \caption{Performance of the MLP-BC model. (a) Profiles of $a_{uv}^*$ predictions of the channel flow at $\Rey_\tau=5200$ by the MLP and MLP-BC models trained on channel flows at $\Rey_\tau=[550, 1000, 2000]$. (b) Profiles of $a_{uv}^*$ predictions of the channel flow at $\Rey_\tau=2000$ by the MLP and MLP-BC models trained on channel flows at $\Rey_\tau=[550, 1000, 5200]$.}
        \label{fig:MLP-BC}
    \end{figure}

    According to Table~\ref{tab:model_performance_r2}, the test $R^2$ scores of the MLP-BC model are higher than the test $R^2$ scores of the MLP model for all four training-test cases, which evidently demonstrates the advantage of enforcing the boundary condition. To gain a more qualitative picture of the model performance, the profiles of the predicted $a_{uv}^*$ for Case 1 and 2 are plotted in Figure~\ref{fig:MLP-BC}. Indeed, the MLP-BC model correctly predicts zero for the $a_{uv}^*$ at the wall, whereas the MLP model predicts non-zero values. 

    Comparing across the four cases, for both models, we observe that in Case 1 and 4 the test $R^2$ is considerably lower than the training $R^2$, while in Case 2 and 3 the test $R^2$ is close to the training $R^2$ (in Case 2 the test $R^2$ is even higher than the training $R^2$). In fact, this pattern exists for other models too. One possible explanation for this is that in Case 1 and 4, the $\Rey_\tau$ of the test set is in a regime not covered by the range of $\Rey_\tau$ in the training set, and so it is harder for the models to generalize.

  \subsection{Performance of the MLP-\texorpdfstring{$\Rey_\tau$ model}{Lg}}
    
    Next, we investigate the performance of the MLP-$\Rey_\tau$ model. In this model, the FCFF network has the same structure as the MLP and the $\Rey_\tau$ is a direct input to the third hidden layer. 
    
\begin{figure}[htp]
        \centering
        \begin{subfigure}{0.49\textwidth}
            \centering
            \includegraphics[width=1.0\columnwidth]{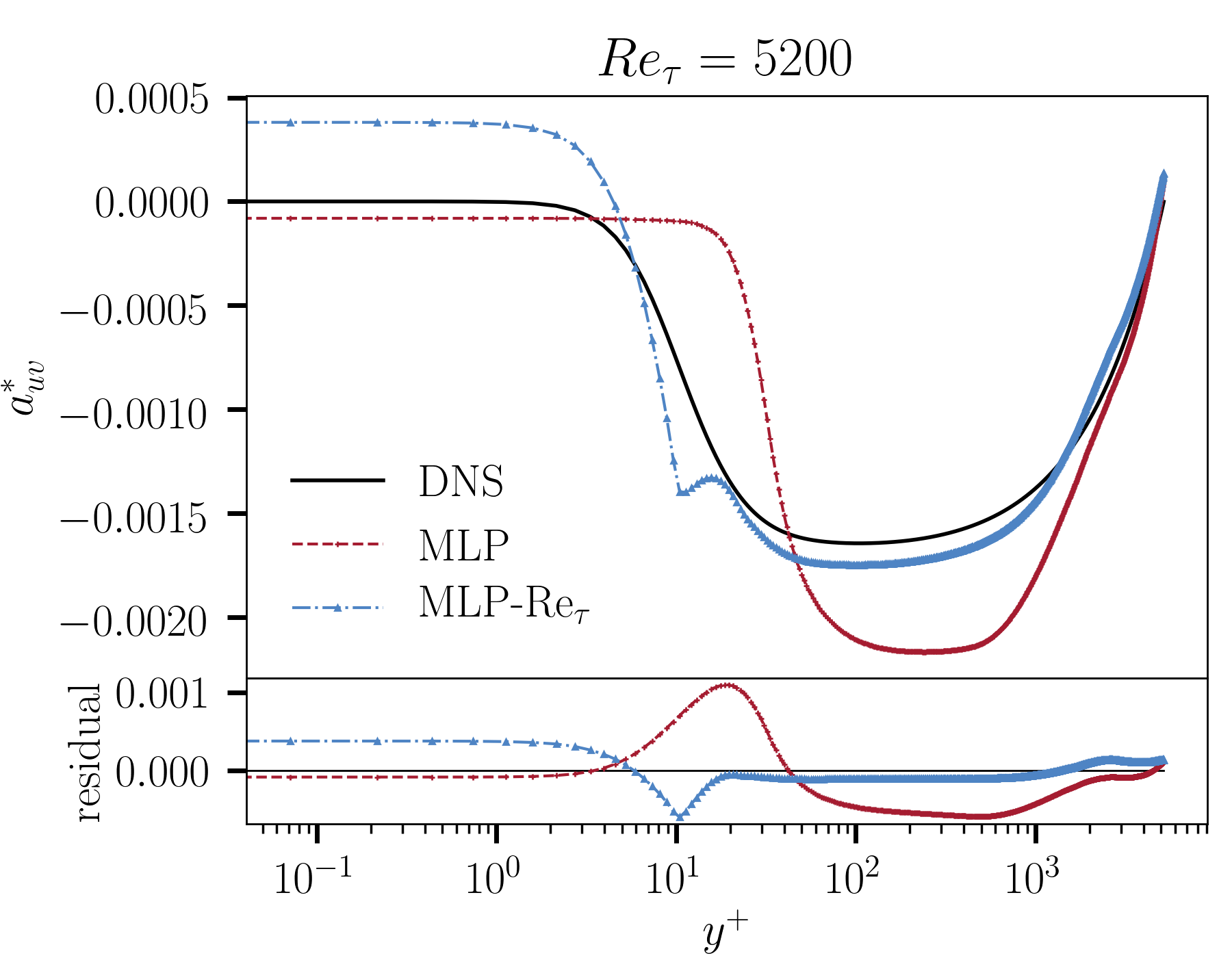}
            \caption{}
        \end{subfigure}
        ~
        \begin{subfigure}{0.49\textwidth}
            \centering
            \includegraphics[width=1.0\columnwidth]{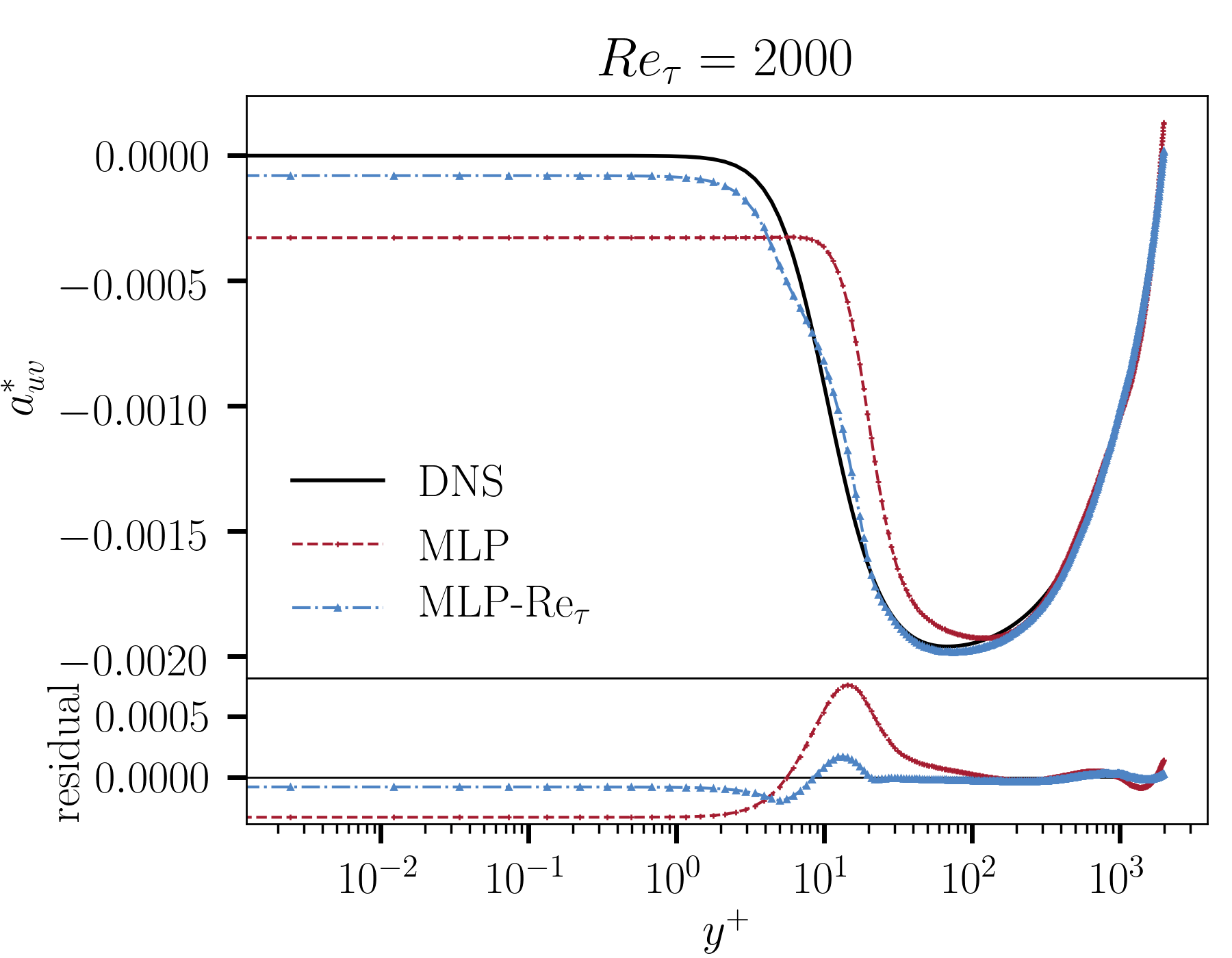}
            \caption{}
        \end{subfigure}
        \caption{Performance of the MLP-$\Rey_\tau$ model. (a) Profiles of $a_{uv}^*$ predictions of the channel flow at $\Rey_\tau=5200$ by the MLP and MLP-$\Rey_\tau$ models trained on channel flows at $\Rey_\tau=[550, 1000, 2000]$. (b) Profiles of $a_{uv}^*$ predictions of the channel flow at $\Rey_\tau=2000$ by the MLP and MLP-$\Rey_\tau$ models trained on channel flows at $\Rey_\tau=[550, 1000, 5200]$.}
        \label{fig:MLP-Retau}
    \end{figure}

    Compared to the MLP, the MLP-$\Rey_\tau$ yields significantly improved predictions: the test $R^2$ scores for the four training-test cases are all above 0.9. The biggest improvement is for Case 1 (test set $Re_{\tau}=5200$) where the $R^2$ raises from 0.5186 to 0.9476.  This demonstrates that Reynolds number injection can greatly enhance the model performance.   Figure~\ref{fig:MLP-Retau} shows the profiles of the predicted $a_{uv}^*$ for Case 1 and 2. In both cases, the MLP-$\Rey_\tau$ provides more accurate predictions than the MLP does in the bulk region of the channel ($y^+>10$).  For $y^{+}<10$, the MLP-$\Rey_\tau$ model does not perform as well as the standard MLP model in Case 1, but is better than the standard MLP model in Case 2.
    
  \subsection{Performance of the MLP-NL model}
    
    In the MLP-NL model, the two FCFF networks both have 5 hidden layers with 50 nodes, the activation function of which is the ELU. The output activation function of the first FCFF network is the sigmoid function. The hyperparameter $\beta$ is 0.1 for enforcing the boundary condition of the fractional derivative order $\alpha$.   
    
\begin{figure}[htp]
        \centering
        \begin{subfigure}{0.49\textwidth}
            \centering
            \includegraphics[width=1.0\columnwidth]{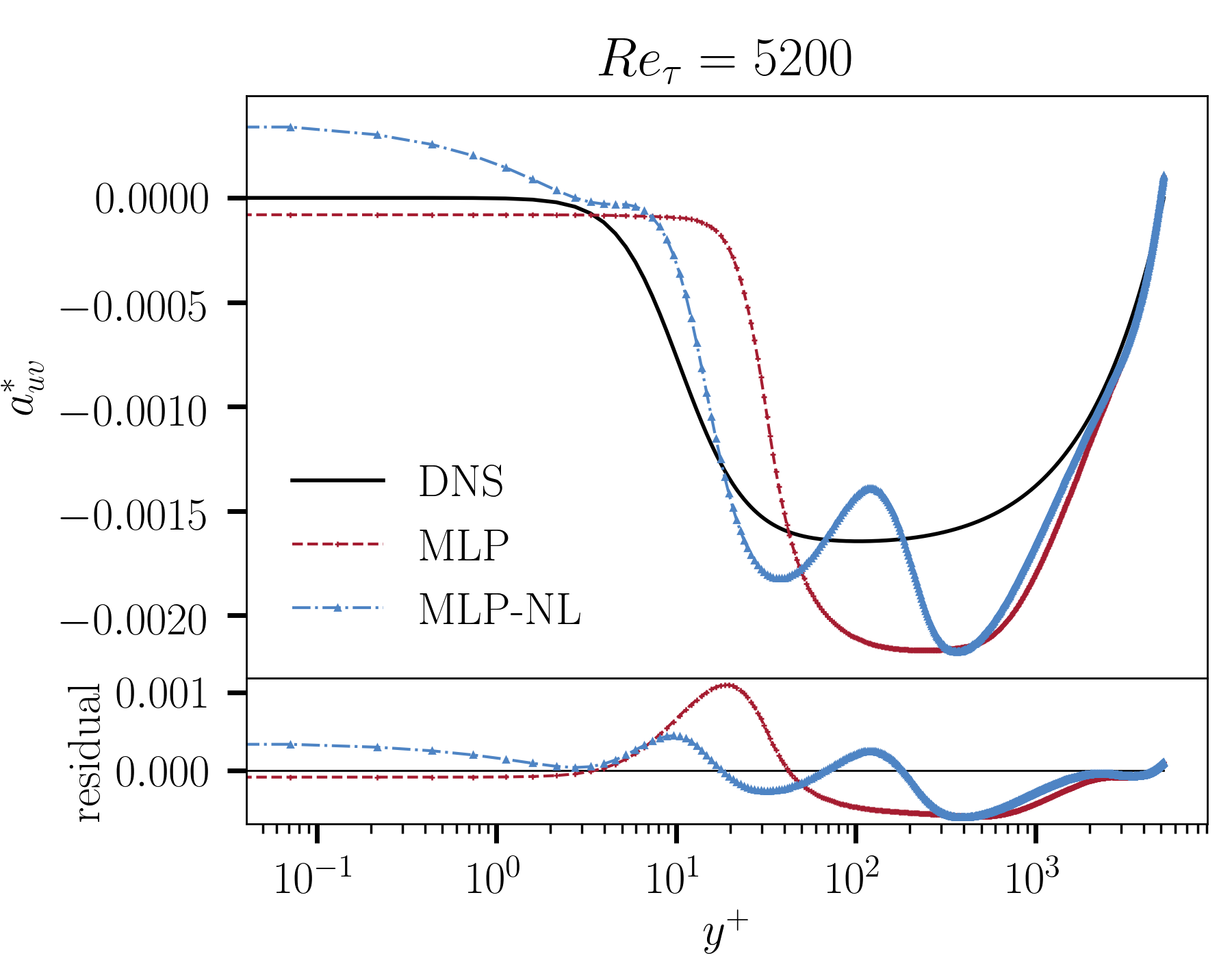}
            \caption{}
        \end{subfigure}
        ~
        \begin{subfigure}{0.49\textwidth}
            \centering
            \includegraphics[width=1.0\columnwidth]{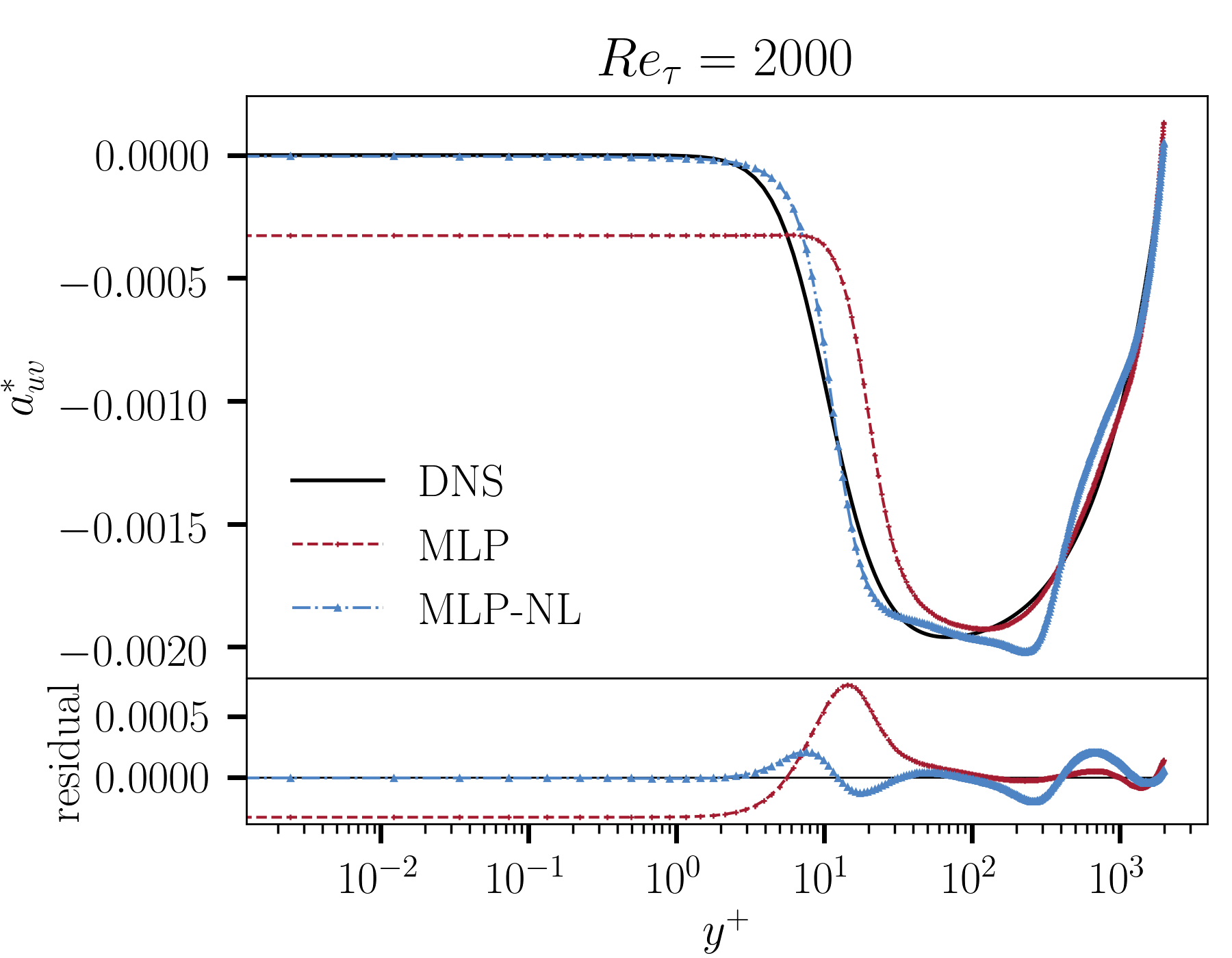}
            \caption{}
        \end{subfigure}
        \caption{Performance of the MLP-NL model. (a) Profiles of $a_{uv}^*$ predictions of the channel flow at $\Rey_\tau=5200$ by the MLP and MLP-NL models trained on channel flows at $\Rey_\tau=[550, 1000, 2000]$. (b) Profiles of $a_{uv}^*$ predictions of the channel flow at $\Rey_\tau=2000$ by the MLP and MLP-NL models trained on channel flows at $\Rey_\tau=[550, 1000, 5200]$.}
        \label{fig:MLP-NL}
    \end{figure}

    Like the previous two models, the MLP-NL model again displays improved performance compared to the MLP. The test $R^2$ scores of the MLP-NL model are in general higher than those of the MLP-BC model, but lower than the MLP-$\Rey_\tau$ model.  Figure~\ref{fig:MLP-NL} shows the profiles of the predicted $a_{uv}^*$ for Case 1 and 2. Noticeably, the predictions of the MLP-NL model in Case 1 are not as good as the predictions in Case 2. As discussed above, the fact that the model struggles with generalizability more in Case 1 is expected because the model is extrapolating to a high $\Rey_\tau$ regime not covered in the training set. 
    
  \subsection{Performance of combined models}
    
    The three modifications have each shown improved performance over the MLP. It is then expected that the combined models would provide more benefits. The MLP-BC-$\Rey_\tau$ model is structured similarly to the MLP-BC and the MLP-$\Rey_\tau$ models: it has got a FCFF network of 5 hidden layers with 50 nodes per layer; the $\Rey_\tau$ is given to the third hidden layer; then the FCFF output is multiplied by a factor $A\lr{y^{+}} = 1-e^{-\beta y^+}$ with $\beta=0.1$ to enforce the boundary condition. Likewise, the non-local combined model MLP-BC-$\Rey_\tau$-NL incorporates the above modifications to the second FCFF network of the MLP-NL model.
    
    Based on Table~\ref{tab:model_performance_r2}, the combined model, MLP-BC-$\Rey_\tau$, indeed achieves higher accuracy than all previous models. However, this is not true for the MLP-BC-$\Rey_\tau$-NL model, the model with supposedly the most capabilities and complexities. In Case 1, the test $R^2$ of the MLP-BC-$\Rey_\tau$-NL model is only 0.8196, smaller than the test $R^2$ of the MLP-$\Rey_\tau$ model, which is 0.9476. In Case 2 and 4, the MLP-BC-$\Rey_\tau$-NL model beats all but the MLP-BC-$\Rey_\tau$ model. Only in Case 3 the MLP-BC-$\Rey_\tau$-NL model gives the best performance. Hence, bringing non-locality into the model through the fractional derivative approach does not always provide additional advantage. 

\begin{figure}[htp]
        \centering
        \begin{subfigure}{0.49\textwidth}
            \centering
            \includegraphics[width=1.0\columnwidth]{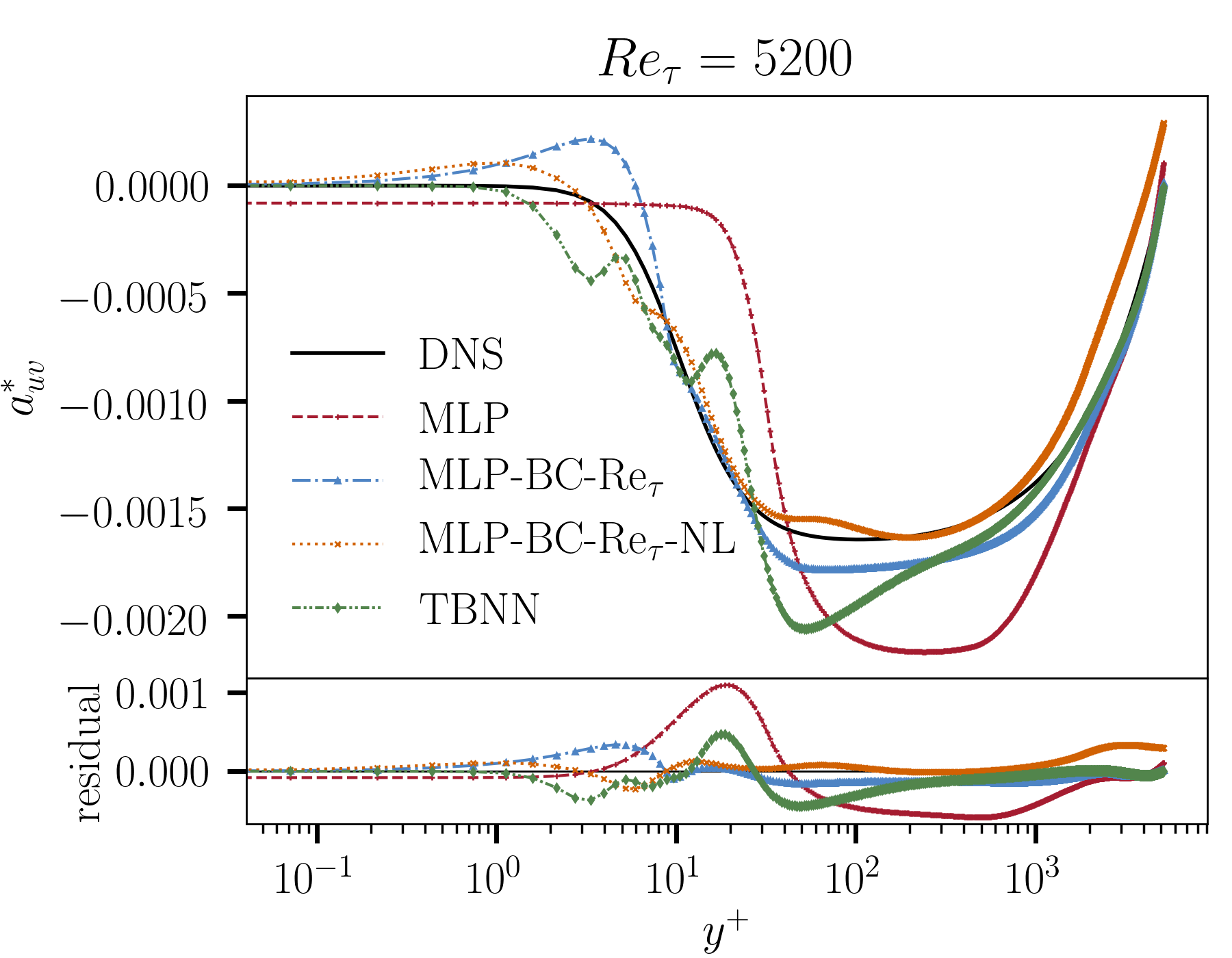}
            \caption{}
        \end{subfigure}
        ~
        \begin{subfigure}{0.49\textwidth}
            \centering
            \includegraphics[width=1.0\columnwidth]{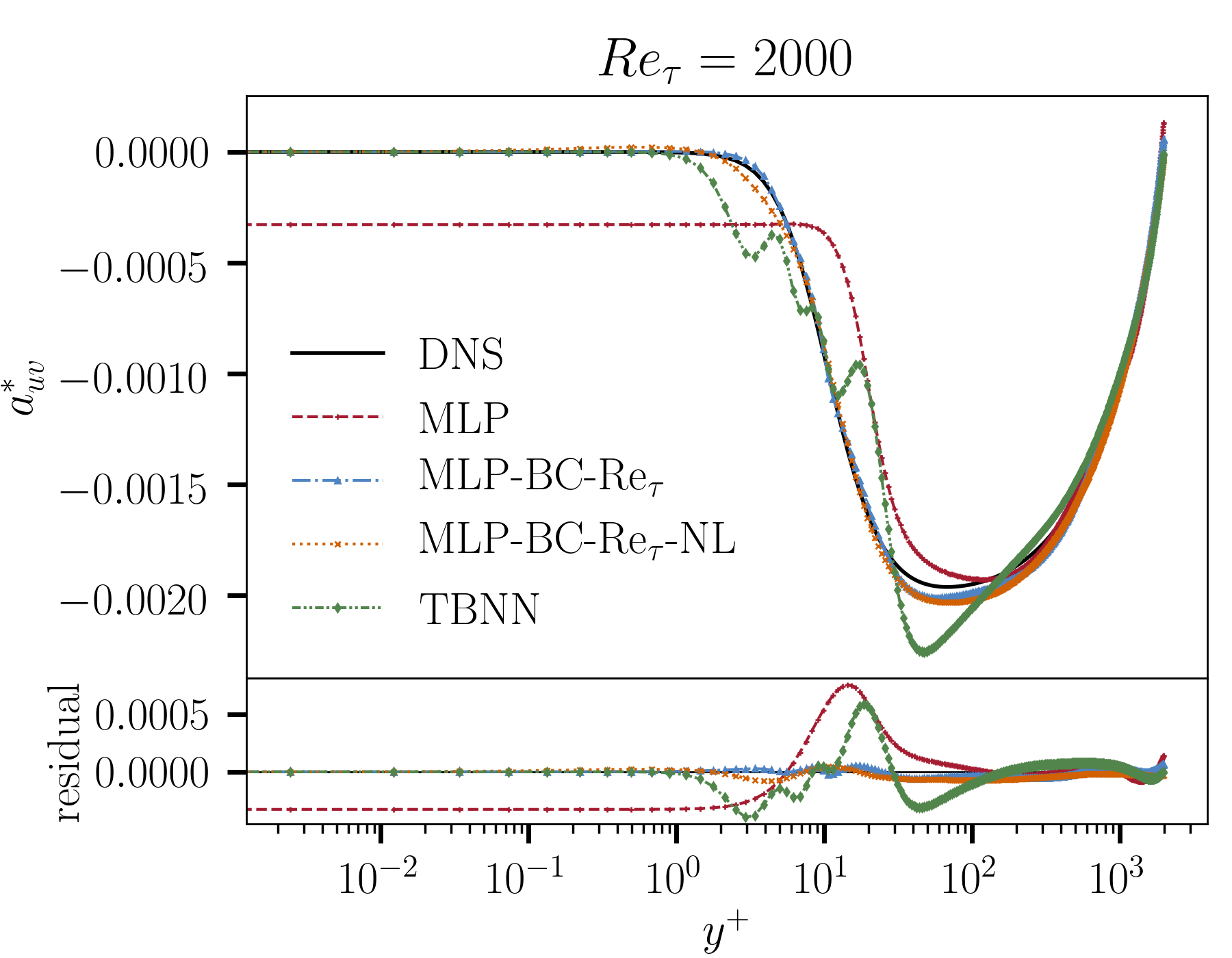}
            \caption{}
        \end{subfigure}
        \caption{Performance of the combined models. (a) Profiles of $a_{uv}^*$ predictions of the channel flow at $\Rey_\tau=5200$ by the MLP, the combined models and the TBNN trained on channel flows at $\Rey_\tau=[550, 1000, 2000]$. (b) Profiles of $a_{uv}^*$ predictions of the channel flow at $\Rey_\tau=2000$ by the MLP, the combined models and the TBNN trained on channel flows at $\Rey_\tau=[550, 1000, 5200]$.}
        \label{fig:combined}
    \end{figure}

    Comparing the proposed models to the TBNN model, we find that the MLP-BC-$\Rey_\tau$ and MLP-$\Rey_\tau$ models consistently outperform the TBNN in all four training-prediction cases. The MLP-BC-$\Rey_\tau$-NL model outperforms the TBNN in Case 2, 3 and 4, yet fails in Case 1. This is verified in the profiles of the $a_{uv}^*$ predictions for Case 1 and 2 in Figure~\ref{fig:combined}. In Case 1, the MLP-BC-$\Rey_\tau$ and the TBNN both yield good predictions, whereas the predictions of the MLP-BC-$\Rey_\tau$-NL model starts deviating from the DNS at about $y^+=1000$. In Case 2, the two combined models and the TBNN all provide good predictions, but the two combined models are significantly better.  This shows that, while for all models it is harder to generalize in Case 1 than in Case 2, different models display varying levels of robustness. The TBNN and the MLP-BC-$\Rey_\tau$ are more robust than the MLP-BC-$\Rey_\tau$-NL.
    

\begin{figure}[htp]
        \centering
        \begin{subfigure}{0.49\textwidth}
            \centering
            \includegraphics[width=1.0\columnwidth]{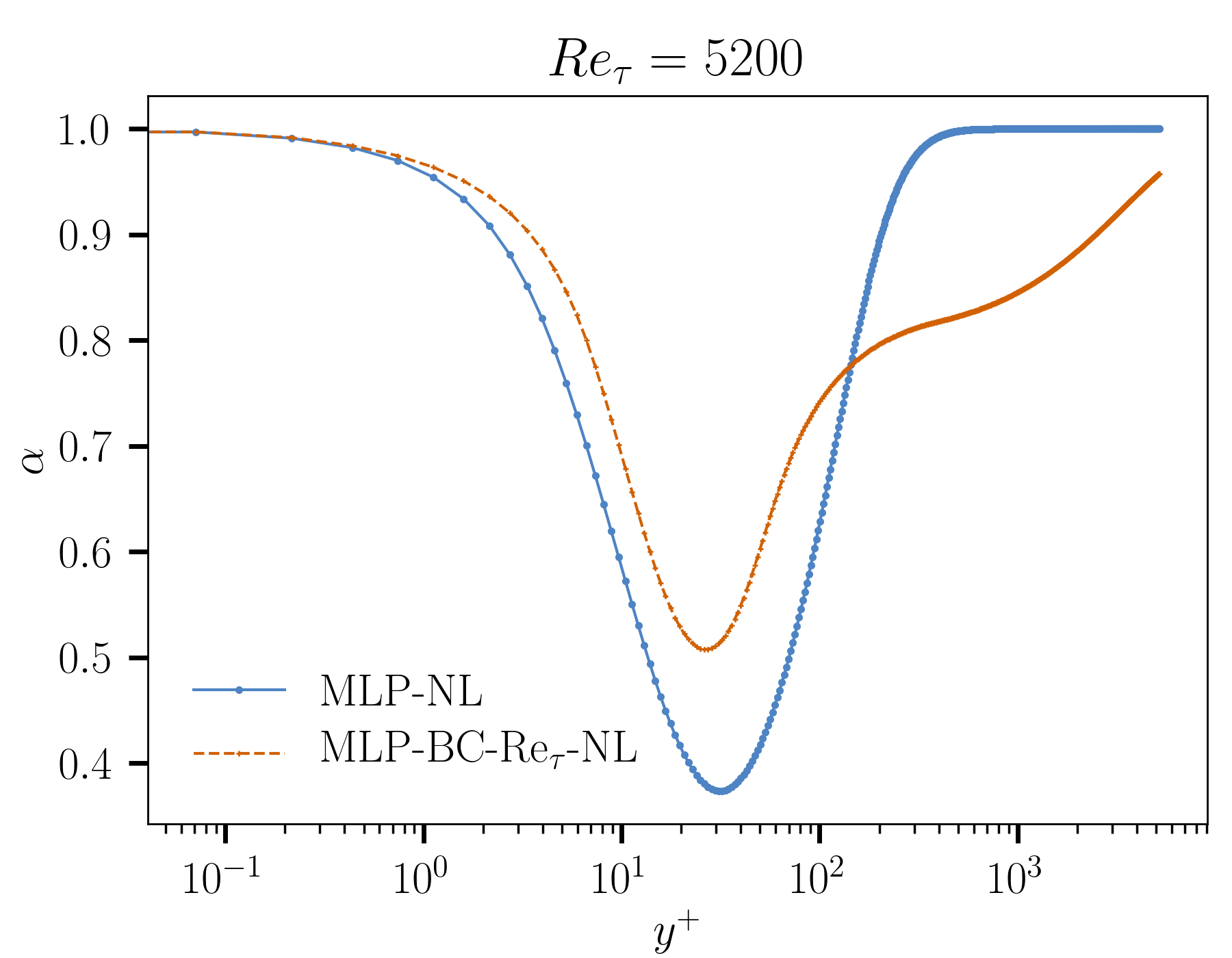}
            \caption{}
        \end{subfigure}
        ~
        \begin{subfigure}{0.49\textwidth}
            \centering
            \includegraphics[width=1.0\columnwidth]{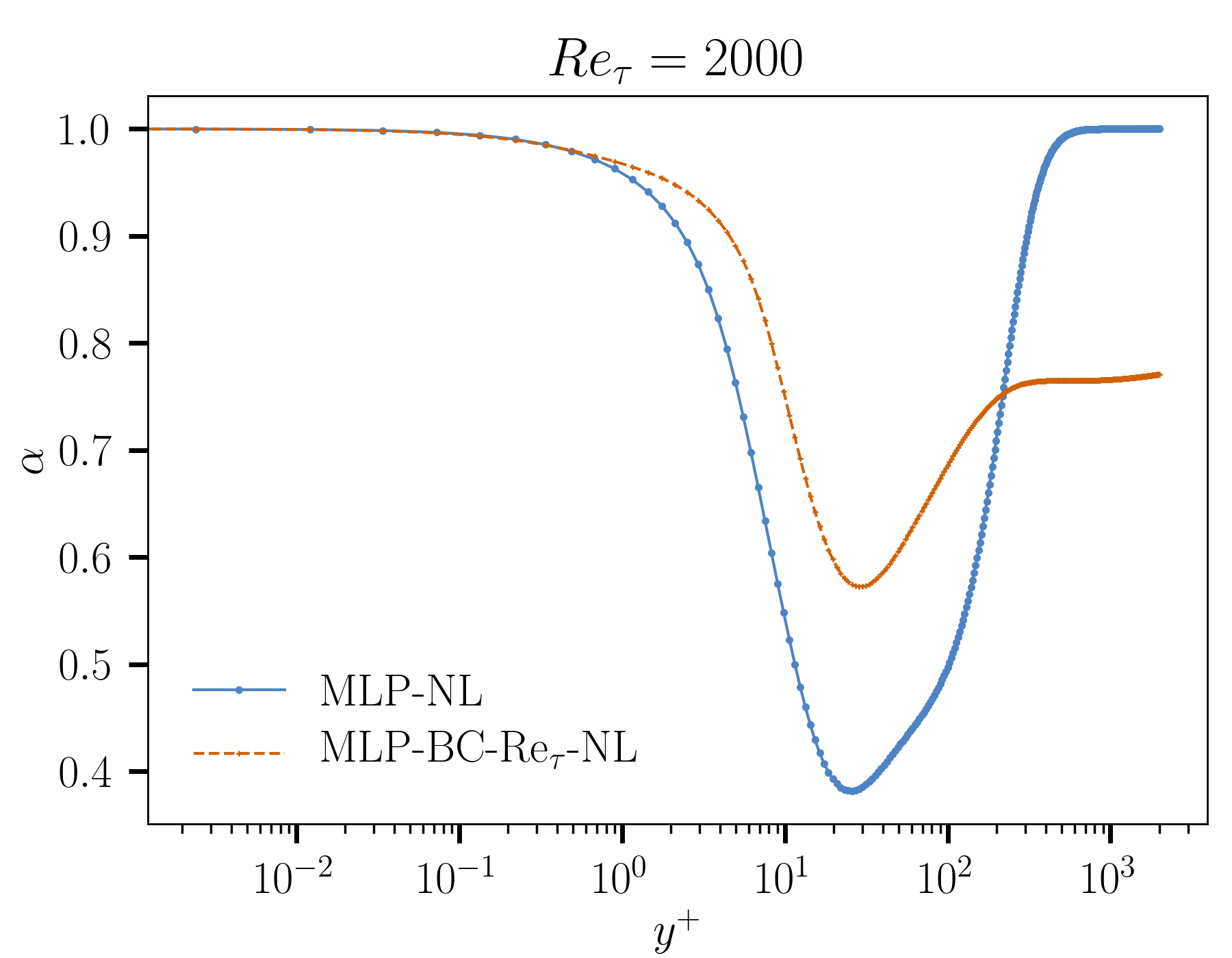}
            \caption{}
        \end{subfigure}
        \caption{Profiles of the fractional derivative order $\alpha$ of the non-local models. (a) The models are trained on channel flows at $\Rey_\tau=[550, 1000, 2000]$ and tested on the channel flow at $\Rey_\tau=5200$. (b) The models are trained on channel flows at $\Rey_\tau=[550, 1000, 5200]$ and tested on the channel flow at $\Rey_\tau=2000$.  Both models show that $\alpha$ is near unity in the immediate vicinity of the wall and reaches its minimum in the near-wall region.  It then increases towards unity again near the bulk region.}
        \label{fig:alpha}
    \end{figure}

    Finally, in Figure~\ref{fig:alpha} we inspect and compare the learned fractional derivative order $\alpha$ of the two non-local models for Case 1 and Case 2. In both cases, and for both models, the $\alpha$ tends to decrease and increase again over the range of $1<y^+<200$. Considering that a bigger deviation from an integer indicates stronger non-local effects, this shows that the non-local effects are more emphasized in specific regions. In particular, the models have learned that non-local effects are significant in the near-wall region and are less significant in the bulk region. Comparing the MLP-NL model and the MLP-BC-$\Rey_\tau$-NL model, we find that the combined model extends the non-local region to large $y^+$ whereas the MLP-NL model learns an $\alpha$ that approaches 1 in the bulk region indicating that the anisotropy tensor is well-modeled by the velocity gradient there. Both models compared in Figure~\ref{fig:alpha} use the fractional derivative to express non-locality.  However, without $Re_{\tau}$ injection, the non-locality is very similar for both Reynolds numbers.  Using Reynolds number injection, the non-locality at different locations acquires roughly the same shape: a dip in the middle before an upturn towards locality near the center of the channel.  The main difference is that the $Re_{\tau}=2000$ case has a smaller $\alpha$ value near the center of the channel than the $Re_{\tau}=5200$ case.  This may be because $Re_{\tau}=2000$ has not reached an asymptotic state yet~\citep{lee2015direct}.  More data at higher Reynolds numbers may provide new insight into this observation.
    
\section{Conclusion}\label{sec:conclusion}

Turbulence modeling with RANS equations is widely used in engineering applications, but properly modeling the Reynolds stress tensor to represent rich and accurate turbulence physics has always been a major challenge.  In the past years, many developments and achievements have been made in modeling the Reynolds stress closure using novel machine learning approaches informed by data, such as deep neural networks. In the present work, we focused on modeling the Reynolds stress closure of the turbulent channel flow with a neural network approach. We proposed three types of modifications to a standard fully-connected network to account for some physical properties of the channel flow: 1.) Reparameterize the network to enforce no-slip boundary condition; 2.) Explicitly provide $\Rey_\tau$ as an input to the network; 3.) Allow for spatial non-locality in the network through a fractional derivative while learning the order of the fractional operator.   To investigate the Reynolds-number generalizability, we designed four training-prediction cases with the channel flow data at four Reynolds numbers. The results showed that compared to the standard network, the three modifications can all improve the predictions, with the most successful results coming from the combination of Reynolds number injection and boundary condition enforcement without the non-local model. We also compared our models to the Tensor Basis Neural Network proposed by~\citet{ling2016reynolds}, which embeds invariance properties into the predicted Reynolds anisotropy tensor. The comparison showed that our best models outperformed the TBNN on this particular flow field. 

There are several directions for future exploration.  A straightforward study to thoroughly explore the hyperparameter space of the models should be performed. For current results, we only performed a very preliminary hyperparameter search. A rigorous hyperparameter optimization would help to validate the present findings. Moreover, in this work we only studied the $u-v$ component of the Reynolds anisotropy tensor for turbulent channel flow. Extensions to other components of the tensor are natural with the current modeling approach and should be investigated in future.  For example, predicting the full Reynolds anisotropy tensor may be interesting.  A direct extension to the current models would allow prediction of the full Reynolds anisotropy tensor while imposing the required symmetry.  More subtle extensions may be possible that take into account the coupling of the components of the tensor.  In addition, the non-local models need to be further examined. Other implementations of non-locality may be possible such as non-local kernel methods.  It is also possible to address non-locality via kinetic formulations in which non-local effects are accessed through local formulations in extended phase-space~\citep{ansumali2004kinetic,succi2018lattice}.  Machine learning approaches in kinetic formulations are still in their infancy and combining the two approaches may provide additional insights.  Beyond these extensions, the models should be tested, modified, and applied to more complex flow fields including fully three-dimensional flows.  Such extensions may involve substantial reformulations of the models.  In particular, in the case of non-local models, the fractional derivative approach will need to be modified.  Another interesting avenue for future exploration is to incorporate uncertainty quantification into the new neural network models.


\renewcommand{\bibname}{References}
\bibliographystyle{ecca}
\bibliography{refs}

\end{document}